\documentclass[%
 reprint,
 superscriptaddress,
 amsmath,amssymb,
 aps,
pra,
]{revtex4-2}

\usepackage{graphicx}
\usepackage{dcolumn}
\usepackage{bm}
\usepackage[normalem]{ulem} 
\usepackage{hyperref}
\hypersetup{colorlinks=true, allcolors=blue} 

\begin{document}

\preprint{APS/123-QED}

\title{Nanocryotron ripple counter integrated with a superconducting nanowire single-photon detector for megapixel arrays
}


\author{Matteo Castellani}
\email{mcaste@mit.edu}

\author{Owen Medeiros}%
\author{Reed A. Foster}%
\author{Alessandro Buzzi}%
\author{Marco Colangelo}%
\affiliation{%
 Electrical Engineering and Computer Science, \\ Massachusetts Institute of Technology, Cambridge, Massachusetts, United States
}%
\author{Joshua C. Bienfang}%
\author{Alessandro Restelli}%
\affiliation{%
Joint Quantum Institute, National Institute of Standards and Technology and University of Maryland, 100 Bureau Drive, Gaithersburg, Maryland 20899, USA}%

\author{Karl K. Berggren}%
\affiliation{%
 Electrical Engineering and Computer Science, \\ Massachusetts Institute of Technology, Cambridge, Massachusetts, United States
}%

\date{May 9, 2024}

\begin{abstract}
Decreasing the number of cables that bring heat into the cryostat is a critical issue for all cryoelectronic devices. Especially, arrays of superconducting nanowire single-photon detectors (SNSPDs) could require more than $10^6$ readout lines. Performing signal processing operations at low temperatures could be a solution. Nanocryotrons, superconducting nanowire three-terminal devices, are good candidates for integrating sensing and electronics on the same technological platform as SNSPDs in photon-counting applications.  
In this work, we demonstrated that it is possible to read out, process, encode, and store the output of SNSPDs using exclusively superconducting nanowires patterned on niobium nitride thin films. 
In particular, we present the design and development of a nanocryotron ripple counter that detects input voltage spikes and converts the number of pulses to an $N$-digit value. The counting base can be tuned from 2 to higher values, enabling higher maximum counts without enlarging the circuit. As a proof-of-principle, we first experimentally demonstrated the building block of the counter, an integer-$N$ frequency divider with $N$ ranging from 2 to 5. Then, we demonstrated photon-counting operations at 405\,nm and 1550\,nm by coupling an SNSPD with a 2-digit nanocryotron counter partially integrated on-chip. The 2-digit counter could operate in either base 2 or base 3 with a bit error rate lower than $2 \times 10^{-4}$ and a count rate of $10^7\,$s$^{-1}$. We simulated circuit architectures for integrated readout of the counter state, and we evaluated the capabilities of reading out an SNSPD megapixel array that would collect up to $10^{12}$ counts per second. The results of this work, combined with our recent publications on a nanocryotron shift register and logic gates, pave the way for the development of nanocryotron processors, from which multiple superconducting platforms may benefit.

\end{abstract}

\maketitle


\section{\label{sec:intro} Introduction}
    Superconducting nanowire single-photon detectors (SNSPDs) have recently been considered for applications such as quantum communication, bio-imaging, and space exploration, thanks to their high efficiency, low jitter, and low dark count rate. In particular, these properties have fueled the development of SNSPD arrays up to the kilopixel scale \cite{esmaeil_zadeh_superconducting_2021, oripov_superconducting_2023}.  A fundamental step to improve resolution and scale for the applications mentioned above is the demonstration of megapixel arrays.
    In such systems, the challenge is to record each individual SNSPD's transient signal, while minimizing the number of output and control lines coming out of the cryostat, and thus the thermal load.
    Different strategies have been explored to reduce the lines using analog multiplexing, such as row-column readout \cite{allman_near-infrared_2015, wollman_kilopixel_2019, allmaras_demonstration_2020}, time multiplexing \cite{zhao_single-photon_2017}, and frequency multiplexing \cite{doerner_frequency-multiplexed_2017}. In these schemes, signal processing and digitization are performed by standard electronics at room temperature. Therefore,  further scaling up would pose limitations to system dimensions and maximum count rate. 
    Low-temperature signal processing and digitization of the output will facilitate the development of megapixel arrays \cite{mccaughan_readout_2018}. Indeed, sensor arrays coupled with rapid-single-flux-quantum (RSFQ) processors have been demonstrated \cite{ miyajima_high-time-resolved_2018, yabuno_scalable_2020}. However, these solutions only encode the pixel address after a detection event, without further processing the output signals.     
    Additionally, Josephson junctions struggle to interface with a high-impedance environment, resulting in a more complicated interface with CMOS readout architectures. Moreover, RSFQ circuits require magnetic shielding \cite{collot_characterization_2016} and have a fabrication process that differs from that of nanowires. This difference has inhibited the integration of these two technologies. 
    An alternative solution could be to use cryo-CMOS readout architectures. However, the system would suffer from the same integrated manufacturing issue as RSFQ circuits and the power dissipation might be prohibitive. 
    Exploiting the same fabrication process and technology for both sensing and electronics would help solve the problems described above, by maximizing on-chip integration.

    A viable solution is to use nanocryotron-based electronics that share the same technological platform as superconducting nanowires. Nanocryotrons (nTrons) \cite{mccaughan_superconducting-nanowire_2014} and heater nanocryotrons (hTrons) \cite{baghdadi_multilayered_2020} are three-terminal devices in which a gate current modulates the critical current of a channel. In nTrons, the gate is electrically coupled to the channel through a small constriction (choke), while hTrons exploit thermal coupling through a heater. 
    The advantages of cryotrons over Josephson junctions are: (1) robustness to magnetic fields \cite{polakovic_superconducting_2020}; (2) possibility to drive high impedances; (3) potential operation at temperatures above 4\,K \cite{merino_two-dimensional_2023}; and (4) high kinetic inductance that reduces inductor footprints. The drawbacks are a 1000$\times$ lower maximum operating frequency and $\sim 100\times$ higher energy consumption. However, in SNSPD systems the detectors would be the bottleneck, by having the same performance limits as nanowires, thus the disadvantage in speed and power is not critically important.
    
    Nanocryotron-based logic gates \cite{buzzi_nanocryotron_2023}, shift registers \cite{foster_superconducting_2023}, encoders \cite{zheng_superconducting_2020}, and memory cells \cite{butters_scalable_2021} have been demonstrated. Moreover, nTrons have been directly coupled to SNSPDs \cite{mccaughan_superconducting-nanowire_2014} and Josephson junctions \cite{zhao_nanocryotron_2017} to pre-amplify the signal before interfacing with semiconductor electronics. Despite the aforementioned achievements, few circuits formed by multiple nTrons that perform a useful and non-trivial logic operation have been demonstrated. 

    A superconducting counter with state storage might be particularly useful in SNSPD arrays because many applications are primarily interested in the number of counts within a given time window. A counter thus enables data reduction, reducing the requirements for carrying analog and digital data off chip.

    In this work, we experimentally demonstrated that the number of photons detected by an SNSPD can be reliably counted, encoded, and stored using a ripple counter based on multiple nTrons. The state of the circuit is encoded in levels of persistent current flowing in superconducting loops, similar to an RSFQ counter \cite{chen_rapid_1999}. 
    An SNSPD coupled to a single superconducting loop storing single-flux quanta to count photons has been demonstrated \cite{onen_single-photon_2020}. 
    A similar approach has been used for optoelectronic spiking neural networks, where Josephson junctions converted SNSPD detection events to circulating currents in a leaky integrating loop \cite{khan_superconducting_2022}. However, the maximum number of stored counts was limited and the output was not digital in both configurations. 
    The circuit proposed in this work can store multi-digit numbers encoded in a base $\geq 2$, without an intrinsic limitation of counts. 
    We designed the ripple counter performing electrothermal simulations and we characterized the bias margins of its building block, an integer-$N$ frequency divider, showing correct counting operations with a tunable $N$. We also characterized the margins of a 2-digit counter operating in bases 2 and 3, and we demonstrated photon counting at 405\,nm and 1550\,nm by coupling an SNSPD to a 2-digit counter fabricated on the same niobium nitride thin film. As a proof-of-concept, they were partially integrated by being interconnected through a resistor surface-mounted on the printed circuit board (PCB).
    
    This proof-of-concept work enables the development of imaging schemes that rely on the high efficiency, wide-wavelength sensitivity, low dark count, and high signal-to-noise ratio of SNSPDs. For some applications, the exact number of detected photons per pixel in a time interval is the critical metric, while the individual photon arrival times are not required. To evaluate the potential of the nanocryotron ripple counter in such applications, we designed and simulated parallel and serial integrated readout schemes based on hTrons. Starting from these simulations, we analyzed design strategies and architectures for reading out an SNSPD megapixel array that can collect tera-counts per second. 

    Digitization and signal processing at low temperatures may also be beneficial in applications other than SNSPD arrays. Indeed, superconducting systems for quantum computing may benefit from a nearby digital technology that is more resistant to flux noise than RSFQ and can easily interface with CMOS. A counter is a critical building block of digital systems. Therefore, the broader goal of this paper is to show that a digital architecture entirely based on superconducting nanowires may be possible and advantageous for scaling up superconducting systems.

\section{\label{sec2:met} Methods}

In this section, we report the methods used to obtain the results described above. In particular, in section \ref{sec2:cur} we describe the design of the ripple counter by showing simulation results. In section \ref{sec2:fab}, we report our fabrication process for detectors and counters. In section \ref{sec2:meas}, we describe the measurement setup and the methods used for the bit-error-rate (BER) estimation of the experimental systems.

 \begin{figure*}[htbp!]
    \centering
    \includegraphics[width=17cm]{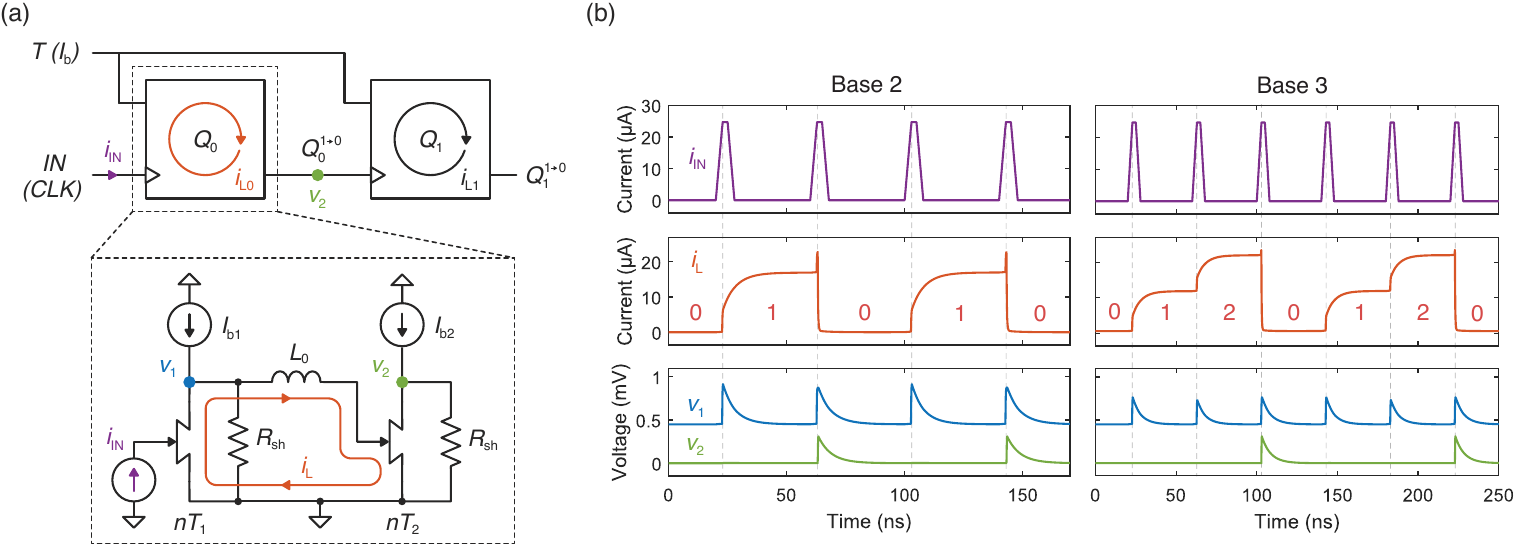}
    \caption{\label{fig: 1bit} Design and simulation of the nanocryotron ripple counter. (a) Behavioral schematic of a 2-digit counter. Each stage operates similarly to a T flip-flop or frequency divider. The inset shows the circuit schematic of one stage. The state $Q$ is encoded in the loop current $i_\text{L}$. The drain inductors of the nTrons are not shown for simplicity. (b) Time-domain simulation of the single stage in base 2 (left) and base 3 (right). The colors associate the waveforms to currents and voltages in the schematic: input current $i_\text{IN}$ (purple, top frame), loop current $i_\text{L}$ (red, middle frame), drain voltage $v_1$ of $nT_1$ (blue, top trace in bottom frame), and drain voltage $v_2$ of $nT_1$ (green, bottom trace in bottom frame). In the middle inset, the digits in red indicate the state $Q$ of the loop. $v_1$ is vertically shifted for clarity. Parameters: $R_\text{sh} = 3$\,$\Omega$, $L_0 = 140$\,nH, $L_\text{d} = 8.5$\,nH, $I^\text{ch}_\text{c} = 190$\,µA, $I^\text{g}_\text{c} = 22.8$\,µA, $I_\text{b1} = 155$\,µA for base 2, $I_\text{b1} = 105$\,µA for base 3, $I_\text{b2} = 115$\,µA.}
    \end{figure*}
    
\subsection{\label{sec2:cur} Circuit design}

    
    The asynchronous nature of ripple counters makes them suitable for counting detection events in imaging applications, in which photon arrival time need not be recorded. A ripple counter can be realized in digital logic with toggle (T) flip-flops in series, each having the toggle always enabled and being clocked by the inverted output of the preceding flip-flop, while the first is clocked asynchronously by the input pulses being counted.
    
    In this work, we demonstrate a ripple counter based on nanocryotrons. Each stage of the superconducting-nanowire-based counter, illustrated in figure \ref{fig: 1bit}\textcolor{blue}{(a)}, operates similarly to a T flip-flop. The output state $Q$ of the flip-flop is encoded in the current $i_\text{L}$ flowing in a superconducting loop. The loop includes two shunted nTrons ($nT_1$ and $nT_2$) and a large kinetic inductor $L_0$. The channels of the two nTrons are current biased ($I_\text{b1}$ and $I_\text{b2}$), and supplying a non-zero bias is analogous to enabling the toggle signal $T$ in a traditional flip-flop. The gate of $nT_1$ receives the input signal (clock). Under correct bias conditions, a current pulse on the input generates a non-superconducting region, called a hotspot, that switches the channel of $nT_1$ and causes most of its bias current to be diverted into its shunt resistor $R_\text{sh}$.  

    A small fraction of the current $\Delta i$ remains in the loop after the $nT_1$ hotspot heals. The presence of $ i_\text{L} = \Delta i$ denotes that $Q = 1$. This current persists in the loop that includes the gate of $nT_2$, and is, by design, lower than the $nT_2$ activation threshold $I^\text{g}_\text{c}$ (critical current of the choke). When a second input pulse arrives, another fraction of the $nT_1$ bias current is diverted into the loop. If $I^\text{g}_\text{c}$ is reached, then a hotspot is generated in $nT_2$, switching its channel. This switching event discharges the loop ($i_\text{L} = 0$; $Q = 0$), resetting the bias current of $nT_1$ to its initial value, and generating an output pulse across the shunt resistor of $nT_2$. After this discharge, the loop is ready to receive an input pulse again.

    In the circuit, there is a small kinetic inductance $L_\text{d} \ll L_0$ on the drain of the nTrons. For simplicity, $L_\text{d}$ is not shown in figure \ref{fig: 1bit}\textcolor{blue}{(a)}. The presence of $L_\text{d}$ and the shunt resistance $R_\text{sh}$ ensures the nTrons do not latch after switching, meaning that their channels always return to a superconducting state. Moreover, $L_\text{d}$ and $R_\text{sh}$ are critical for setting the amount of current injected into the loop and the maximum allowed input (clock) frequency.

    We designed the circuit using LTspice simulations \footnote{The identification of any commercial product or trade name does not imply endorsement or recommendation by the National Institute of Standards and Technology.} based on models of nanocryotrons \cite{castellani_design_2020}, which rely on an electrothermal model of a superconducting nanowire \cite{berggren_superconducting_2018}. 
    The left graph in figure \ref{fig: 1bit}\textcolor{blue}{(b)} shows a simulation of the circuit described above. In particular, the top inset shows the input current pulses generated by an ideal current source. The middle inset shows the current stored in the loop with associated state $Q$:  $i_\text{L} = 0$\,µA, 18\,µA is equivalent to $Q = 0, \ 1$ respectively. The bottom inset shows the spikes generated by the nTrons when a current is injected or removed from the loop. A voltage pulse on $nT_1$ ($v_1$) indicates a positive edge of the $Q$ state ($Q = 0 \rightarrow  Q = 1$). A voltage pulse on $nT_2$ ($v_2$) indicates a negative edge on the $Q$ state ($Q = 1 \rightarrow  Q = 0$). Therefore, using $v_2$ ($Q^{1\rightarrow 0}$) to clock the next flip-flop, as shown in figure \ref{fig: 1bit}\textcolor{blue}{(a)}, is analogous to clocking with $\overline{Q}$ in a standard ripple counter.

    The single stage of the counter differs from a standard T flip-flop because the toggle port (bias lines of the nTrons) receives analog values. In particular, the value of $I_\text{b1}$ sets the amount of current $\Delta i$ injected into the gate of $nT_2$. 
    Therefore, decreasing $I_\text{b1}$ increases the number of possible current states allowed in the loop before $nT_2$ switches. 
    This characteristic allows tuning of the base $N$ in which $Q$ is encoded, where $N$ is the number of input pulses needed to make $nT_2$ switch. 
    In other words, the T flip-flop becomes an integer-$N$ frequency divider and it is possible to create a multi-digit ripple counter in base $N$. 
    To do so the bias currents of all the nTrons must be equally decreased. 
    The right graph of figure \ref{fig: 1bit}\textcolor{blue}{(b)} shows that the circuit can count in base 3: $i_\text{L} = 0$\,µA, $11$\,µA, $22$\,µA is associated with $Q = 0,\ 1,\ 2$ respectively.

    Lowering $I_\text{b1}$ such that $nT_1$ of the first stage does not switch when stimulated is equivalent to setting $T$ to zero. The circuit stops counting and its state remains stored. 
    By slightly modifying the circuit of the single stage, it is possible to read out and reset the state of the counter upon the application of an external $READ$ signal. Alternatively, a serial readout scheme can be implemented by making the counter act like a shift register. The simulations of these architectures are described in section \ref{sec2:read}.

\begin{figure*}[htbp!]
    \includegraphics[width=15.5cm]{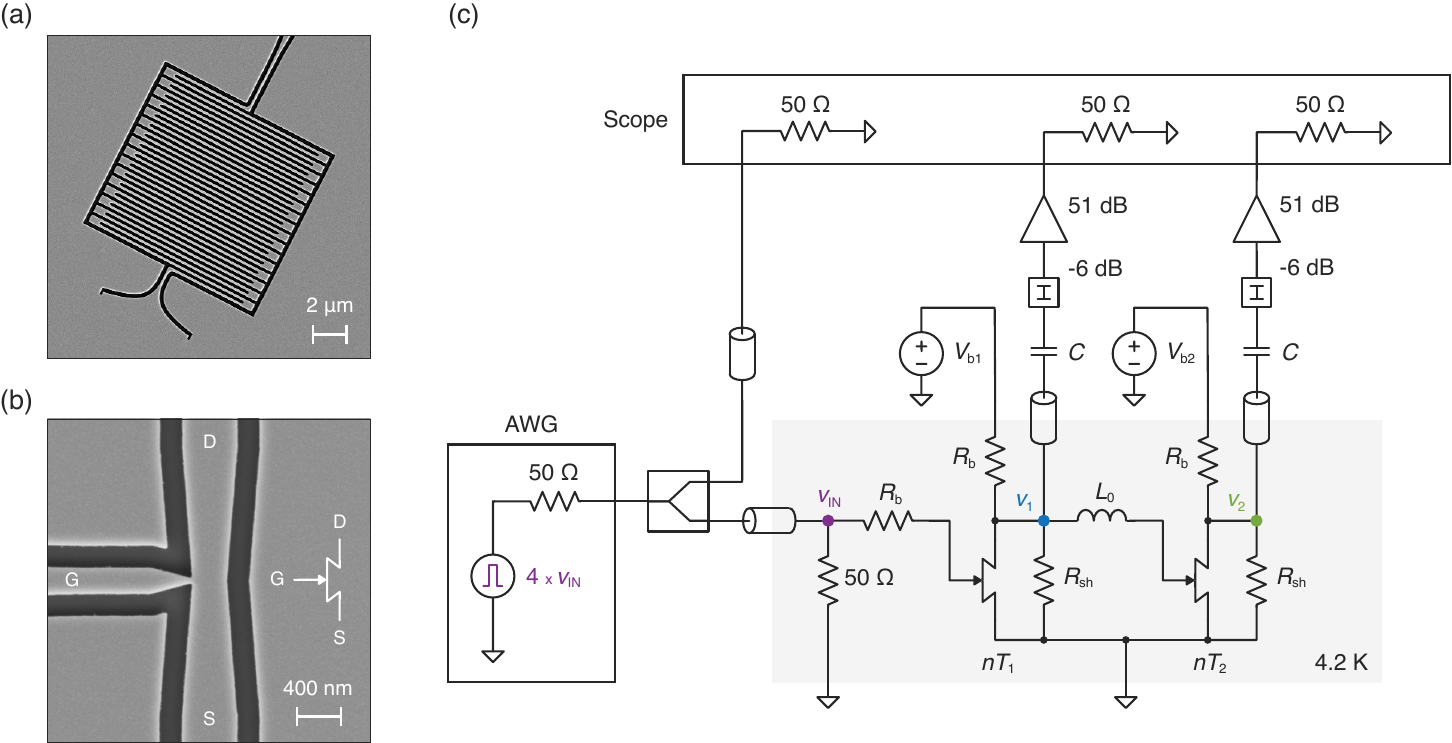}
    \caption{Fabricated devices and measurement setup. (a) Scanning electron micrograph of a 200\,nm wide  SNSPD occupying a $12\,\text{µm} \times 14\,\text{µm}$ area. The gray regions are niobium nitride, and the dark borders of the traces show the underlying SiO$_2$ substrate. The ground plane surrounds all the traces. (b)  Scanning electron micrograph of a nanocryotron (nTron) with the associated circuit symbol. The device has three ports: drain (D), source (S), and gate (G). The choke is the 50\,nm wide constriction that connects the gate to the 300\,nm wide channel. Results shown in the next sections were obtained using a 40\,nm wide choke and a 250\,nm wide channel. (c) Setup for the time-domain measurement of a single stage of the counter. The input signal, generated by an arbitrary waveform generator (AWG) ($4\times v_\text{IN}$), is split and read by the oscilloscope as well as being sent to the device ($v_\text{IN}$). Coaxial cables (50\,$\Omega$ impedance) connect the device from 4.2\,K to room temperature electronics. A 50\,$\Omega$ resistor in front of the input bias resistor ensures matching between the PCB and the coaxial cable. The components enclosed in the gray box are on the PCB at 4.2\,K. The nTrons are biased with room-temperature batteries ($V_\text{b1}$, $V_\text{b2}$) and low-temperature resistors $R_\text{b}$. Each output signal passes through a bias tee ($C$) with the DC line left open, an attenuator, and two amplifiers (providing a total gain of 45 dB), before going to the oscilloscope. The impedance of the scope is set to 50\,$\Omega$. \label{fig: setup} }
    \end{figure*}

\subsection{\label{sec2:fab} Fabrication process}

    We fabricated four $1\,\text{cm} \times 1\,\text{cm}$ chips having three different types of device: (1) single stages of the counter; (2) 2-digit counters; and (3) SNSPDs.
    For all of them, we varied some key parameters (e.g. loop inductance, nTron gate width) to find the operational geometries. Correct counting operations were observed in the devices of all four chips. 
    We fabricated two samples with multiple stand-alone stages (SPG613, SPG687), and two other chips, both having 2-digit counters and SNSPDs (SPG716, SPG781).
    
    The fabrication process was the same for all the samples \cite{castellani_design_2020}: we deposited an 18\,nm thick niobium nitride (NbN) film on a 300\,nm thick thermal-silicon-oxide (SiO$_2$) on silicon substrate, with an AJA sputtering system \cite{Note1} at 160\,W of peak power. We spun the positive-tone resist ZEP-530A \cite{Note1} at 524\,rad/s (5000 rpm) for 60\,s, and baked it at 180\,$^{\circ}$C for 120\,s. We defined the device geometries using electron-beam lithography (Elionix ELS-F125 \cite{Note1}). Then, we developed the resist in O-Xylene at 5\,$^{\circ}$C for 60\,s and rinsed the samples in isopropanol (IPA) for 30\,s. Afterward, we patterned the devices through reactive ion etching (RIE) based on CF$_4$ at 50\,W, and rinsed the chips in N-Methyl-2-pyrrolidone (NMP) at 70\,$^{\circ}$C for 1 hr.
    
    The kinetic inductance was estimated to be 20\,pH per square, according to results of previous work \cite{castellani_design_2020}. The inductors were realized with meandered nanowires to minimize their footprint. The values of inductance for devices used here are: $L_0 \approx 140\,$nH, $L_\text{d} \approx 10\,$nH. The meandered nanowires composing the inductors were 400\,nm wide. 
    The results in section \ref{sec2:snspd} were obtained using a 200\,nm wide SNSPD occupying an area of $12\,\text{µm} \times 14\,\text{µm}$, and having a fill factor of 50\,\% (see figure \ref{fig: setup}\textcolor{blue}{(a)}). The nTrons had a 40\,nm wide choke, and a 250\,nm wide channel. Figure \ref{fig: setup}\textcolor{blue}{(b)} shows an example of nTron geometry. 

    As a proof-of-principle, we patterned all the devices on 18\,nm films to minimize the footprint of the counter and simplify its characterization without focusing on optimizing the performance of SNSPDs. Indeed, using film thicknesses on the order of 5\,nm would be optimal to obtain detectors with saturated photon detection efficiency but the footprint of counters would increase (wider wires) to ensure the same critical currents. Having high critical currents, and thus larger signals, helps improve the signal-to-noise ratio and margins, facilitating the characterization. 
    In section \ref{sec:dis}, we propose a multi-layer structure to obtain good performance for both the detector and the counter.

\subsection{\label{sec2:meas} Measurement methods}
    In this section, we introduce the methods used to evaluate the device's performance and we describe the measurement setup.
    Firstly, we characterized stand-alone stages, then 2-digit counters, both being stimulated by current pulses through arbitrary waveform generators (AWGs). Finally, we tested 2-digit counters controlled by SNSPD pulses.
    
    For each circuit, we estimated the bit error rate (BER) at different bias points to find the bias range over which accurate counting was possible in each base (bias margins). The estimation was done by applying streams of current pulses (ranging in length from $10^2$ to $10^4$) to the input of the devices. The use of more input pulses was limited by the testing apparatus, thus only BER as low as $10^{-4}$ were confirmed experimentally. 
    We did not directly measure the state $Q_\text{n}$ of each loop but we could observe the state transitions $Q^{1 \rightarrow 0}_\text{n}$. From the latter, we extrapolated the BER, defined as the number of bit errors over the total number of input pulses. To find a bit error, we observed the spikes on the $Q^{1 \rightarrow 0}_\text{n}$ lines generated in response to each input pulse. When we saw a spike, the bit $Q^{1 \rightarrow 0}_\text{n}$ was set to 1, otherwise $Q^{1 \rightarrow 0}_\text{n}$ was set to 0. We compared one by one, and in chronological order, the bits of each $Q^{1 \rightarrow 0}_\text{n}$ line with the elements of a reference vector of expected values. If a discrepancy was found, the subsequent reference bits were updated to avoid multiple counts of the same error. A bit error in one stage was defined as follows: (1) $Q^{1 \rightarrow 0}_\text{n}$ was 1 instead of 0, meaning that the loop current was higher than expected so $Q_\text{n}$ reset to 0. Therefore, at the next input pulse, a spike should have not been observed; or (2) $Q^{1 \rightarrow 0}_\text{n}$ was 0 instead of 1, meaning that the loop current was not high enough to make the second nTron switch. In this case, $Q^{1 \rightarrow 0}_\text{n}$ should have spiked at the next clock cycle.
    If an error occurred in one stage of a multi-digit counter, the next reference bits of all the loops were updated to avoid counting the same error on multiple stages.

    We performed all the measurements in a liquid-helium Dewar at 4.2\,K. 
    Each chip was glued with GE Varnish \cite{Note1} to a PCB, and the devices were connected to the PCB through aluminum wire bonds. The PCB was then placed on the cold head of a custom cryogenic probe with 28 spring-loaded RF sub-miniature push-on connectors \cite{butters_digital_2022}.
    An optical fiber was installed on the cold head of the probe at a height of about 2\,cm from the PCB to allow characterizations with flood laser illumination. Moreover, a superconducting solenoid was attached to the cold head to study the effects of an out-of-plane magnetic field on the bias margins of the devices.  

 \begin{figure*}[htbp!]
    \centering
    \includegraphics[width=16.5cm]{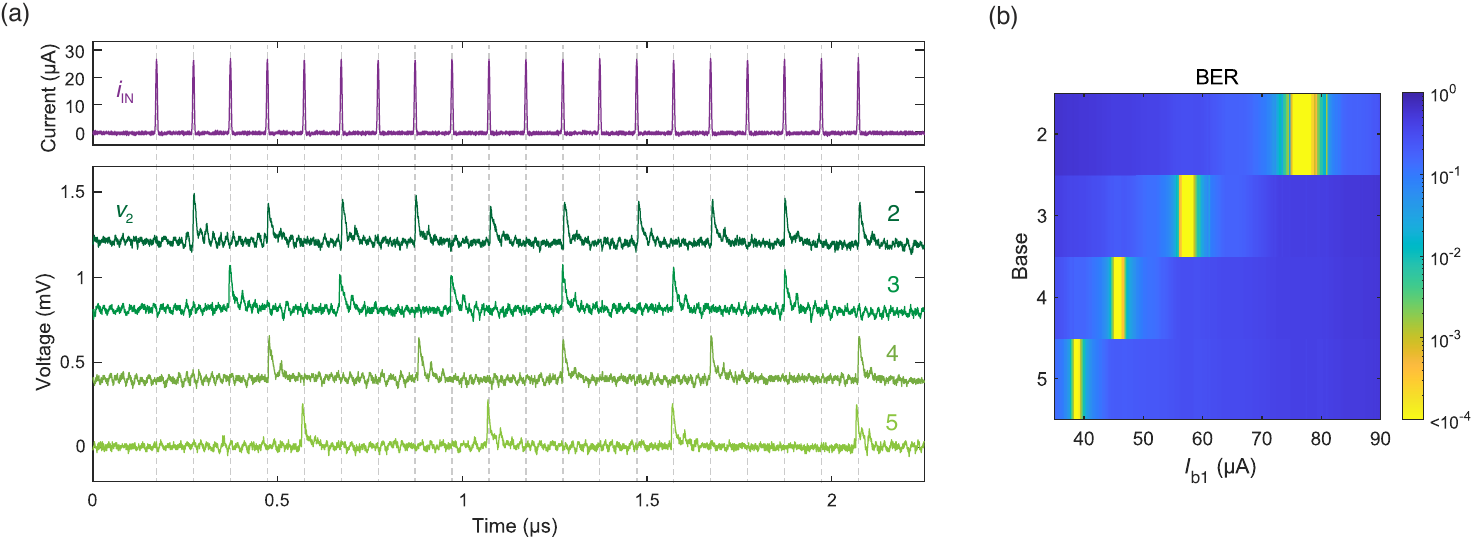}
    \caption{\label{fig: 1bit_meas} Characterization of a single stage. (a) Experimental time-domain behavior of one stage as an integer-$N$ frequency divider: input current $i_\text{IN}$ (purple, top), and drain voltage $v_2$ of $nT_2$ (shades of green, bottom) after amplification. The voltage amplitude is divided by the gain of the amplifiers at 1\,GHz reported in the datasheet (45 dB). The numbers on the right indicate $N$. The waveforms are vertically shifted for clarity. (b) Experimental bit error rate (BER) of the stage as a function of the bias current $I_\text{b1}$ for different bases, with a 5\,MHz input frequency (clock). $I_\text{b2} = 85 $\,µA and $i^\text{max}_\text{IN} = 72$\,µA (input peak current) were kept constant. The two figures were obtained from two different devices: C32 on SPG613; and C01 on SPG687 respectively. “The two figures were obtained from two different devices: C32 on SPG613; and C01 on SPG687 respectively. The two device geometries were almost identical (nTrons in C32 had a longer path to ground than nTrons in C01, resulting in a 10\,\% larger drain-to-ground inductance in C32). The resolution of the captured traces used for figure (b) was lower to speed up the sweep.}
    \end{figure*}
    
    We characterized single stages of the counter (results in section \ref{sec2:char}) in the cryogenic probe, using the setup shown in figure \ref{fig: setup}\textcolor{blue}{(c)}.
    The bias currents were provided by low-noise batteries (SRS SIM928 \cite{Note1}) in series with 10\,k$\Omega$ resistors ($R_\text{b}$) to obtain nearly ideal current sources. 87\,pF surface-mount bypass capacitors were used on all the bias lines at 4.2\,K.
    The nTrons were shunted with $3\,\Omega$ resistors ($R_\text{sh}$).
    All the bias resistors, shunts, and capacitors were surface mounted on the PCB at 4.2\,K to limit the thermal noise. 
    The input voltage signal was generated by an AWG (Agilent AWG33622A \cite{Note1}), and read out through a splitter, before being sent to the PCB. Matching conditions between the AWG and the PCB were ensured by a 50\,$\Omega$ resistor to ground, placed on the input pin of the device. The input voltage was converted to a current signal with a 10\,k$\Omega$ series resistor, before reaching the input port of the counter.
    The input signal was a burst stream of 8\,ns wide pulses of varying lengths.
    The AC part of each of the two output signals ($Q^{1 \rightarrow 0}_\text{0}$ and $Q^{1 \rightarrow 0}_\text{1}$) was filtered by a bias tee and attenuated by 6 dB to limit reflections. Then, it passed through two low-noise amplifiers (RF Bay LNA-2500, bandwidth: 10\,kHz to 2500\,MHz, gain: 25\,dB, and RF Bay LNA-2000, bandwidth: 10\,kHz to 2000\,MHz, gain: 26\,dB \cite{Note1}), before being read by a 2 GHz, real-time oscilloscope (LeCroy 620Zi \cite{Note1}). The internal 20\,MHz low-pass filter of the scope was applied.

    We tested the 2-digit counters (results in section \ref{sec2:2bitc}) with a setup similar to the one described above. 
    The only differences were: (1) The number of output lines was 3 instead of 2; (2) for preliminary characterizations we used Agilent AWG33622A and the scope, but for the bias-margins estimation we used a programmable AWG (Keysight M3202A \cite{Note1}), and a data acquisition system (DAQ) (Keysight M3102A \cite{Note1}) to automate and speed up the measurements. For the bias-margins estimation, the input signal was a pseudo-random stream of 5\,ns wide pulses.

    After characterizing the 2-digit counter alone, we coupled it to an SNSPD on the same chip.
    The detector was biased by a battery in series with a 10\,k$\Omega$ resistor ($R_\text{bd}$), and shunted with $R_\text{d}=50\,\Omega$ to avoid latching. The SNSPD was connected to the input of the counter through a 1\,$\Omega$ resistor ($R_\text{s}$) to avoid trapped flux in the loop formed by the gate of $nT_1$ and the detector.
    The current diverted from the SNSPD to the counter was maximized by $R_\text{s}$ being much smaller than $R_\text{d}$. 
    All the resistors were surface mounted on the PCB at 4\,K. In particular, $R_\text{s}$ had one terminal wire bonded to the SNSPD and the other terminal wire bonded to the input port of the counter (wire-bond lengths were typically a few mm). 
    The counter was biased and read out with the same setup used for the preliminary characterizations of the stand-alone 2-digit counter.
    
    We flood illuminated the entire chip through the fiber installed on the cold head. The fiber was aligned to the center of the die, where the SNSPD was patterned, to maximize the optical coupling. 
    The results shown in section \ref{sec2:snspd} were obtained with a 405\,nm continuous laser nominally operating at 60\,µW. The correct functioning of the system was demonstrated also with a 1550\,nm continuous laser set to a nominal power of 9.9\,mW (with a 150\,nm wide SNSPD, area: $10\,\text{µm}\times 10\,\text{µm}$, fill factor: 45\,\%) which showed similar performance. The continuous illumination made the photon sensor generate a random stream of voltage pulses that we measured by adding an output line (bias tee, attenuator, amplifier, and scope) on the shunt of the SNSPD. Therefore, we did not use the AWG for the BER estimation of this system. For this measurement, the internal 200\,MHz low-pass filter of the scope was applied.

\section{\label{sec:res} Experimental results}
    In this section, we show our experimental results for the three characterized systems: (1) a stand-alone stage of the counter in section \ref{sec2:char}; (2) a 2-digit counter in section \ref{sec2:2bitc}; and (3) a 2-digit counter coupled to an SNSPD in section \ref{sec2:snspd}.
    We also describe readout schemes for the state stored in the counter, by showing simulation results in section \ref{sec2:read}. Moreover, we propose a possible architecture for an SNSPD megapixel array that relies on nanocryotron counters, in section \ref{sec:array}.  

\subsection{\label{sec2:char} Characterization of a single stage}
    \begin{figure*}[htbp!]
    \includegraphics[width=17.6cm]{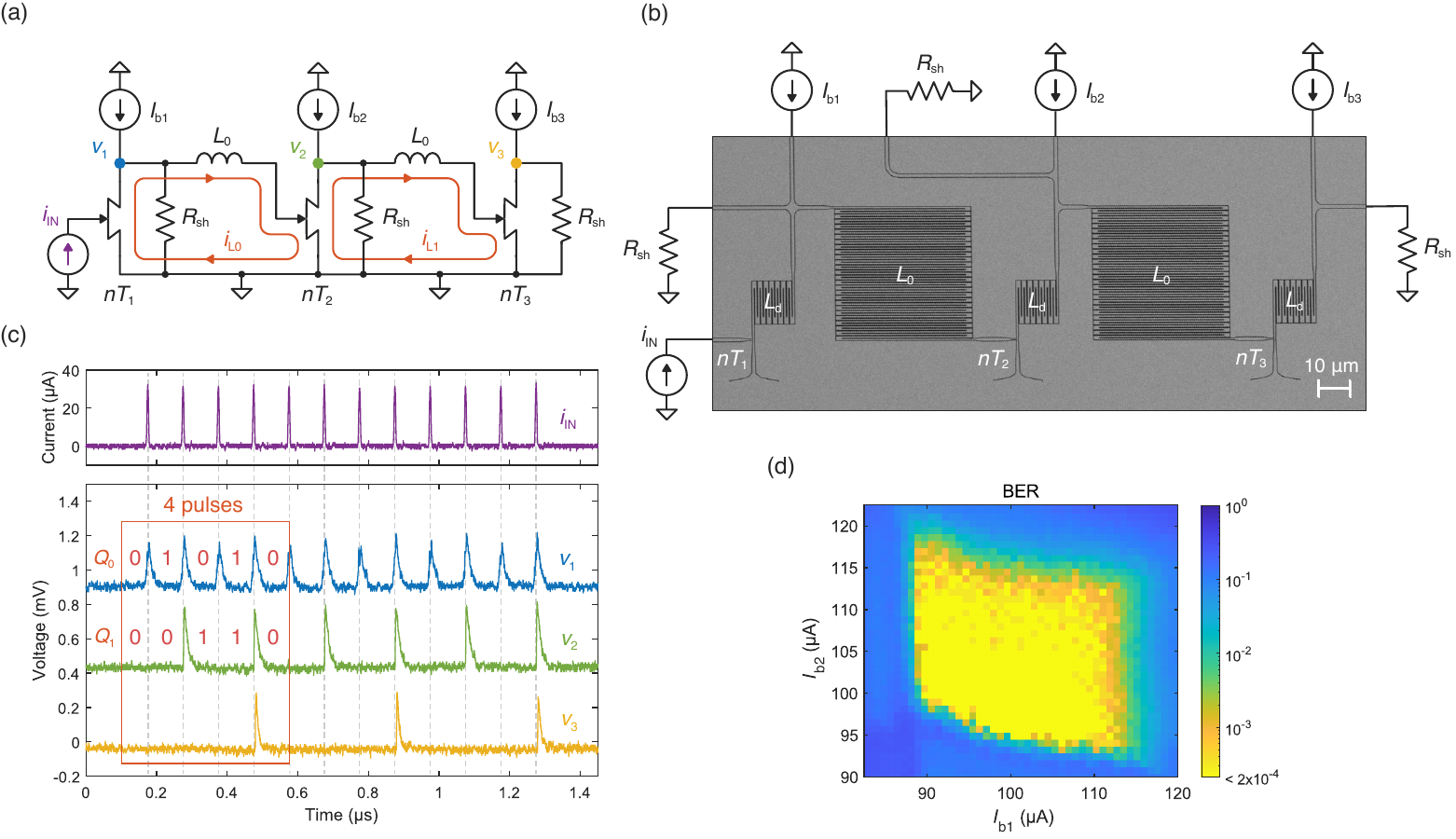}
    \caption{\label{fig: 2bit} Characterization of a superconducting nanowire 2-digit counter. (a) Circuit schematic of a 2-digit superconducting nanowire counter. The least significant digit $Q_0$ was encoded in $i_\text{L0}$, the most significant digit $Q_1$ in $i_\text{L1}$. The central nTron resets the first stage and sets the second stage. The drain inductors of the nTrons are not shown for simplicity. (b) Scanning electron micrograph of the 2-digit counter with external components. The resistors are surface mounted on the PCB, the current sources are implemented by voltage sources with 10\,k$\Omega$ resistors in series. The gray regions are niobium nitride, and the dark borders of the traces show the underlying SiO$_2$ substrate. The ground plane surrounds all the traces. (c) Experimental time-domain behavior. Colors and labels associate the waveforms with currents and voltages in the schematic. The voltage traces were acquired after the amplification. In the figures, voltage amplitudes are divided by the gain of the amplifiers at 1\,GHz reported in the datasheet (45 dB) The red box highlights the count of 4 pulses, showing the digits $Q_0$ and $Q_1$. The waveforms are vertically shifted for clarity. (d) Experimental bit error rate (BER) as a function of $I_\text{b1}$ and $I_\text{b2}$ in base 2, with a pseudo-random asynchronous input. The minimum delay between adjacent pulses was 100\,ns. $I_\text{b3}$ and $I_\text{IN, max}$ (input peak current) were kept constant. Figures (b), (c), and (d) were obtained from the same device on SPG716.}
    \end{figure*}

\begin{figure*}[htbp!]
    \includegraphics[width=16cm]{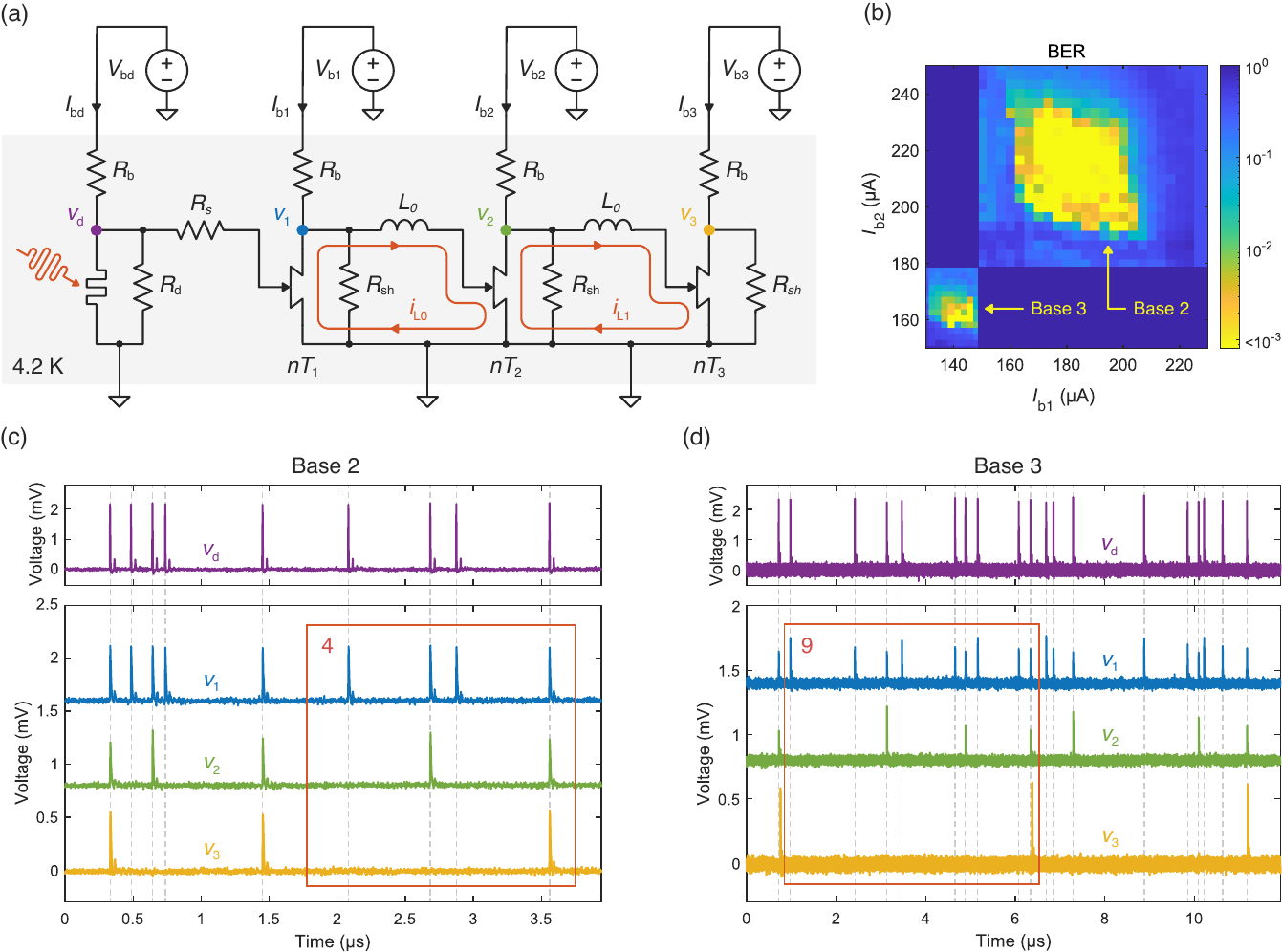}
    \caption{\label{fig: 2bit_SNSPD} Characterization of a superconducting nanowire 2-digit counter coupled to an SNSPD flood illuminated at 405\,nm. (a) Circuit schematic of the tested structure. The SNSPD was coupled to the $nT_1$ through $R_\text{s}$.  (b) Experimental bit error rate (BER) as a function of $I_\text{b1}$  and $I_\text{b2}$, with the signal from SNSPD as input. $I_\text{b3}$ and $I_\text{bd}$ were kept constant. The system can operate in base 2 or base 3. Regions with BER = 1 (dark blue) were not analyzed, as the counter should not operate there. (c) Experimental time-domain behavior in base 2. (d) Experimental time-domain behavior in base 3. For both plots, colors, and labels associate the waveforms with currents and voltages in the schematic. The voltage traces were acquired after the amplification. In the figures, voltage amplitudes are divided by the gain of the amplifiers at 1\,GHz reported in the datasheet (45 dB). The red boxes highlight the count of 4 or 9 pulses. The waveforms are vertically shifted for clarity. Figures b, c, and d were obtained from the same device on SPG781.}
    \end{figure*}

    Before demonstrating a multi-digit counter, we characterized a stand-alone stage to show its behavior as a T flip-flop or integer-$N$ frequency divider. Moreover, we confirmed that the same device could be used to count in different bases $N$ by tuning the bias current of $nT_1$. 
    Figure \ref{fig: 1bit_meas}\textcolor{blue}{(a)} shows the AWG-generated input pulses (bursts of 20 pulses with a 100\,ns pulse period and a 100\,µs burst period) and the associated output voltage $v_2$ of $nT_2$ for different bases. The bias current of $nT_2$ was kept constant. 
    The periodic spiking behavior of $v_2$ was consistent with the simulation results in figure \ref{fig: 1bit}\textcolor{blue}{(b)}: in base 2, the spiking frequency of $v_2$ was half of the clock frequency; in base $N$, the spiking frequency was divided by $N$; $v_1$ followed the clock signal (not shown for simplicity).
    Therefore, we confirmed that the stage behaved as expected without directly measuring the state $Q$.

    Figure \ref{fig: 1bit_meas}\textcolor{blue}{(b)} shows the experimental bias margins of the loop for four bases $N = 2, 3, 4, 5$, with BER that can be less than $10^{-4}$ using a 5\,MHz input signal (clock). A lower bound for the BER may be found by using more than $10^{4}$ incoming pulses. In base 6, at $I_\text{b1} = 20$\,µA (not shown in the figure), we observed a BER of $10^{-3}$.
    We distinguished the four different ranges of operation by decreasing $I_\text{b1}$ from $80$\,µA to $35$\,µA. As expected, the lower $I_\text{b1}$, the higher the base. However, the transition between regions at low BER and different $N$ is not sharp. It passes through values of BER as high as $10^{-1}$ for all the bases. There are two possible reasons why this behavior could occur. Ideally, during each period of $v_2$, the current injected into the loop by $nT_1$ and the current removed by $nT_2$ are the same. In simulations, we observed that the two currents can be unbalanced by a small amount in the regions with high BER. These small amounts can add up to $i_\text{L}$ after a certain number of input pulses, altering the state of the loop periodically. Therefore, a periodic pattern that differs from the one in figure \ref{fig: 1bit_meas}\textcolor{blue}{(a)} could be obtained. 
   The second cause of errors may be electrical noise, which generates non-periodic random bit errors. This may happen when the loop is in its highest state and the value of $i_\text{L}$ is very close to the critical current $I^\text{g}_\text{c}$ of the second gate. This effect occurs more often for higher bases because $\Delta i$ is smaller.
    In figure \ref{fig: 1bit_meas}\textcolor{blue}{(b)}, the dimensions of the lighter (yellow) regions and the spacing between them decrease by lowering $I_\text{b1}$ because $\Delta i$ decreases. As a consequence, at lower bases, the bias current had to be reduced by a smaller amount to pass from base $N$ to $N+1$. In base 2, the optimal bias point was 76.5\,µA with a $\pm 2\,\%$ variation.

    The circuit could correctly operate in base 2 with a 10\,MHz input signal in the presence of an external out-of-plane magnetic field up to 3\,mT. At 3\,mT, the minimum bit error rate was $10^{-4}$. More information can be found in the supplemental material \footnote{See Supplemental Material at [URL will be inserted by publisher]}.

    From the experimental transient voltages on the nanocryotrons, we estimated the energy consumption of each nTron to be approximately 0.5\,fJ per spike. The energy dissipated in the bias resistors and the cooling cost was not included in this calculation.

\subsection{\label{sec2:2bitc} Characterization of a 2-digit counter}
    A multi-digit counter is formed by a cascade of stages. To simplify the architecture by minimizing the number of nanocryotrons, the second nTron of each loop can be used as the first of the subsequent loop. The simplification is possible because the negative edge of $Q_\text{n}$ must generate a positive edge on $Q_\text{n+1}$. So $v_2$ of the first stage is equivalent to $v_1$ of the second stage. 
    Figure \ref{fig: 2bit}\textcolor{blue}{(a)} shows the circuit schematic of a 2-digit counter implemented with this simplification. 
    Figure \ref{fig: 2bit}\textcolor{blue}{(b)} shows a scanning electron micrograph of the 2-digit counter, with illustrated external resistors and current sources. 
    Figure \ref{fig: 2bit}\textcolor{blue}{(c)} shows the experimental time-domain behavior in base 2 when bursts of 12 pulses with a period of 100\,µs are applied. As expected, each stage divides the spiking frequency of the preceding one, suggesting that the system is correctly counting in modulo 4; it resets to the zero state every 4 input pulses. The voltage $v_3$ acts as an overflow signal; other stages can be connected in cascade to increase the number of digits. 

    To study the bias margins of a 2-digit counter it is necessary to sweep the bias currents of two nTrons, because in general,  $\Delta i$ may be different for each loop. Figure \ref{fig: 2bit}\textcolor{blue}{(d)} shows the BER for different values of $I_\text{b1}$ and $I_\text{b2}$, where $I_\text{b3}$ was kept constant. The minimum time interval between two pulses in the input pseudo-random stream was 100\,ns. 
    
    The errors in the two loops were correlated; this fact was taken into account in the estimation of the bit error rate. 
    The optimal bias region was approximately centered around $(I_\text{b1}, I_\text{b2}) = (100, 105)$ µA, with a $\pm 10\,\%$ margin.
    The shape of this region was not a square because $I_\text{b1}$ and $I_\text{b2}$ were not completely equivalent in the circuit. This lack of equivalency was because the first nTron was stimulated by the input signal that did not vary during the sweep. The gate of the second nTron was activated by the current in the first loop, which depended on $I_\text{b1}$. By decreasing the gate current, the suppression of the critical current in the channel was smaller. Therefore, for lower values of $I_\text{b1}$, higher values of $I_\text{b2}$ were needed to make the counter operate properly.
    The right and top edges of the correct-bias region were more affected by random bit errors, resulting in a wider grey zone in these regions. We observed that these errors almost disappeared when we lowered the average input pulse rate to $10^{6}$\,s$^{-1}$.
    For pulse rates higher than $10^{7}$\,s$^{-1}$, the bias margins progressively shrank. \cite{foster_superconducting_2023} presented a detailed experimental analysis of this effect in a nTron shift register, whose structure is similar to the one of the counter.   

\subsection{\label{sec2:snspd}  Characterization of a 2-digit counter coupled to an SNSPD}
    
    We demonstrated photon counting operations coupling an SNSPD to a 2-digit counter. Figure \ref{fig: 2bit_SNSPD}\textcolor{blue}{(a)} shows the circuit schematic of the system.
    The detection efficiency of the SNSPD did not saturate either at 1550\,nm or at 405\,nm but single-photon detection was confirmed (results of the detector's characterization are shown in the supplemental material \footnotemark[\value{footnote}]. Therefore, the detector was sufficient for demonstrating counting of all the detected photons. 
    The SNSPD was biased close to its switching current to maximize the photon sensitivity and the output signal. The dark count rate was negligible with respect to the rate of input pulses.
    We designed the gate and channel of $nT_1$ to be about $20\,\%$ narrower than those of the other nTrons in the circuit. This difference made the first nTron more sensitive to low input currents from the SNSPD so that the counting base could be finely tuned without sacrificing the sensitivity of the gate.
    
    Figure \ref{fig: 2bit_SNSPD}\textcolor{blue}{(c)} shows the experimental time-domain behavior of the system in base 2. The voltage $v_\text{d}$ on the SNSPD was a random stream of pulses, which were correctly detected by the first nTron. The same considerations of section \ref{sec2:2bitc} are valid for the second inset in the image: the circuit counts modulo 4 (\textit{i.e.}, base 2) since each stage divides the spiking frequency of the preceding one. 
    In Figure \ref{fig: 2bit_SNSPD}\textcolor{blue}{(d)}, the same system counts modulo 9 (\textit{i.e.}, base 3). In both time-domain plots, the distribution of peak heights in $v_1$ and $v_2$ suggests that $N-1$ different current levels were progressively stored in the loops: the peak voltage of the spike was proportional to the channel current, which was lowered by a $\Delta i$ after each input pulse before the subsequent nTron fired.
    
    Figure \ref{fig: 2bit_SNSPD}\textcolor{blue}{(b)} shows the bias margins for the system in bases 2 and 3, with a minimum BER lower than $10^{-3}$. Lower values were not verified for this case due to the long measurement times required (the rate of detection events of the SNSPD was low) and the limited speed of the equipment used (the fast programmable AWG and DAQ from Keysight were not available at the time of the measurement). The bias current of the SNSPD and $I_\text{b3}$ were kept constant. The square-like shape of the low-BER region for base 2 was consistent with that of the stand-alone 2-digit counter in figure \ref{fig: 2bit}\textcolor{blue}{(d)}. Moreover, the base-2 region was approximately centered around $(I_\text{b1}, I_\text{b2}) = (185, 215)$\,µA, with a $\pm 10\,\%$ margin. The values differed from the previous section because the critical currents of this device were higher.
    The low resolution of the current sweep limited our ability to analyze the BER margins in base 3. 
    In this experiment, an additional possible cause of bit errors may have been photon-generated hotspots somewhere in the counter (e.g. in the nTron gates).

\begin{figure*}[htbp!]
    \includegraphics[width=15cm]{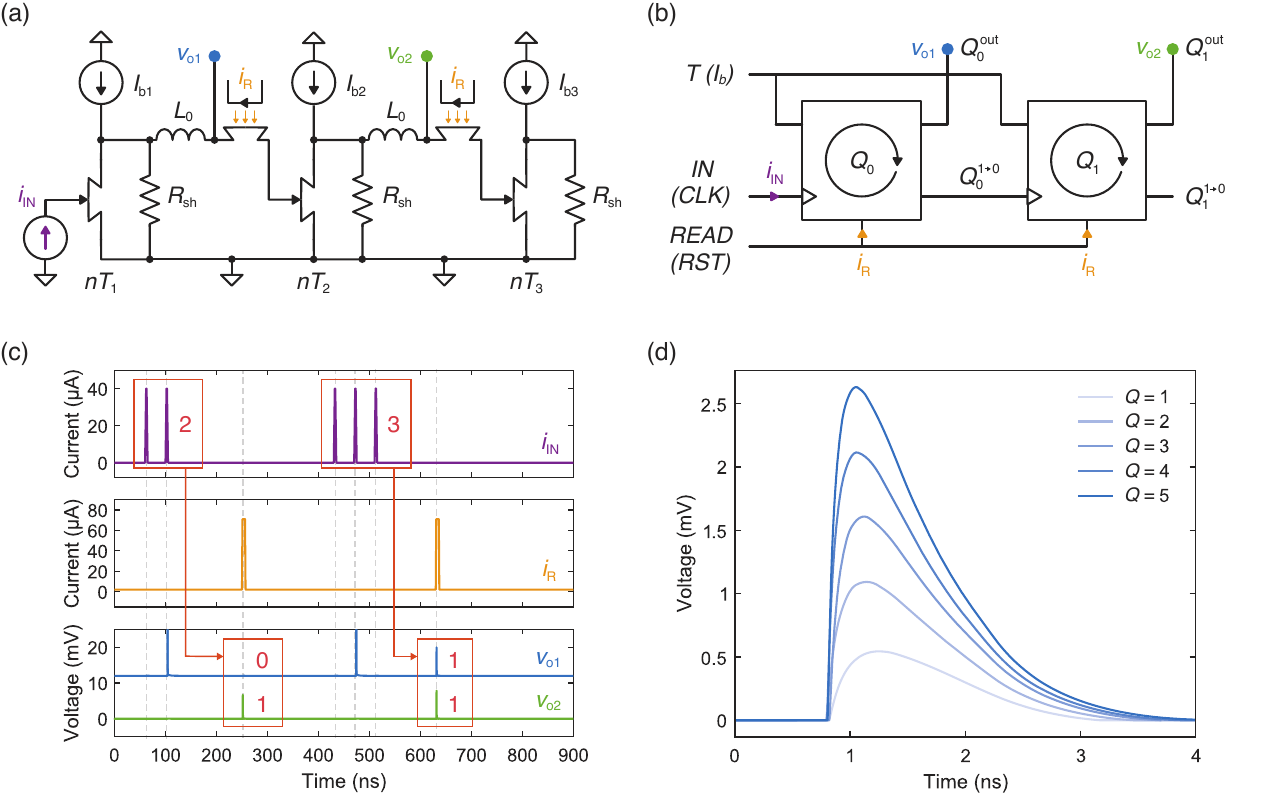}
    \caption{\label{fig: read} Design and simulation of a parallel-readout method for a superconducting nanowire counter. (a) Circuit schematic of a 2-digit counter. An hTron is introduced in each loop to generate a hotspot upon the application of a heater current $i_\text{R}$. The hotspots create voltage pulses on the outputs $v_\text{o1}$ and $v_\text{o2}$ and reset the loops. (b) Behavioral schematic of the 2-digit counter. Adding the hTrons is equivalent to introducing a $READ$ signal and an output line $Q^\text{out}_\text{n}$ to each stage. The $READ$ port also acts as a $RESET$ ($RST$). (c) Simulated time-domain behavior of the 2-digit counter with parallel readout in base 2. Two input pulses are counted and read out. Afterward, the counter is reset and three input pulses are counted and read out. In base 2, the state of the counter is encoded in the absence or presence of spikes on $Q^\text{out}_\text{n}$. The red boxes show the digital state on the output lines. The two pulses on $v_\text{o1}$ generated during the counting interval are not read out. The circuit has been simulated also with 0, 1, 4, and 5 input pulses but the waveforms are not shown here for simplicity. The colors and labels associate the waveforms with currents and voltages in the schematic. (d) Simulated output voltage of a 1-digit counter working in base 6, with parallel readout configuration, for different numbers of input pulses. The loop is stimulated with 1, 2, 3, 4, or 5 current pulses, then the state is read out. The load impedance on the output line is 50\,$\Omega$. In the simulation, the states 0 through 5 are distinguished by 0.5\,mV.}
    \end{figure*}

    \section{\label{sec:res_sim} Other results}

    \subsection{\label{sec2:read} Simulation of counter readout schemes }
        
    In the previous sections, we characterized the counters by measuring in real time the $Q^{1 \rightarrow 0}_\text{n}$ signals. 
    However, in imaging applications, a common mode of operation is to accumulate counts, then disable the counter and read out the states $Q_\text{n}$ when required. In this section, we propose circuit architectures that can implement this procedure. 
    
    Figure \ref{fig: read}\textcolor{blue}{(a)} shows the circuit schematic of a 2-digit counter that allows parallel readout of the digits, and figure \ref{fig: read}\textcolor{blue}{(b)} shows the corresponding behavioral schematic. 
    When a $READ$ pulse is applied, the states $Q_\text{n}$ of the stages are simultaneously read out on the $Q^\text{out}_\text{n}$ ports, and automatically reset. 
    This configuration is obtained by placing a heater nanocryotron (hTron) in each loop. In particular, the superconducting channel of the hTron is connected in series between the loop inductor and the gate of $nT_2$. The heater can be a normal metal resistor or a superconducting nanowire in close proximity to the channel. The simulation results shown here were obtained using a model of a multi-layer hTron with a normal metal heater \cite{castellani_design_2020}.
    A $READ$ current pulse is simultaneously sent to all the heaters so that all the hTron channels switch to the normal state and the loop currents are diverted to the output ports. Therefore, voltage pulses proportional to $i_\text{Ln}$ are generated on the output loads at the same time. After the measurement, the hTrons cool down and the circuit is reset for the next counting operations. 
    
    Figure \ref{fig: read}\textcolor{blue}{(c)} shows the simulated time-domain behavior of the circuit in base 2. It can count two pulses, store the digital number, and return the associated bits in parallel when the READ pulse is activated. Afterward, the circuit repeats the procedure with three input pulses. In base 2, the bits are encoded in the presence (value 1) or absence (value 0) of voltage pulses on the readout lines. 
    
    In higher bases, the loop currents are encoded in the amplitude of the pulses on the readout lines. The simulation result in figure \ref{fig: read}\textcolor{blue}{(d)} shows that it is possible to distinguish between 5 different levels of circulating current by measuring the voltage at the output node of a single loop (the zero level is not shown). The voltage step between adjacent levels is about 0.5\,mV, a value that can be resolved by an analog-to-digital converter (ADC). 

    The counter can also be designed to act as a shift register, similar to \cite{foster_superconducting_2023}, for applications that require a serial rather than parallel output.
    In the supplemental material \footnotemark[\value{footnote}], we show simulation results for a 2-digit counter read out as a shift register. 
    This scheme reduces the number of output lines and does not require hTrons, but necessitates a different clock signal for each stage to shift the state out. The output is a stream of pulses encoding the digital number stored in the counter. This operation can only be implemented in base 2.

\subsection{\label{sec:array} Design of a readout scheme for SNSPD megapixel arrays}

    \begin{figure*}[htbp!]
    \includegraphics[width=17cm]{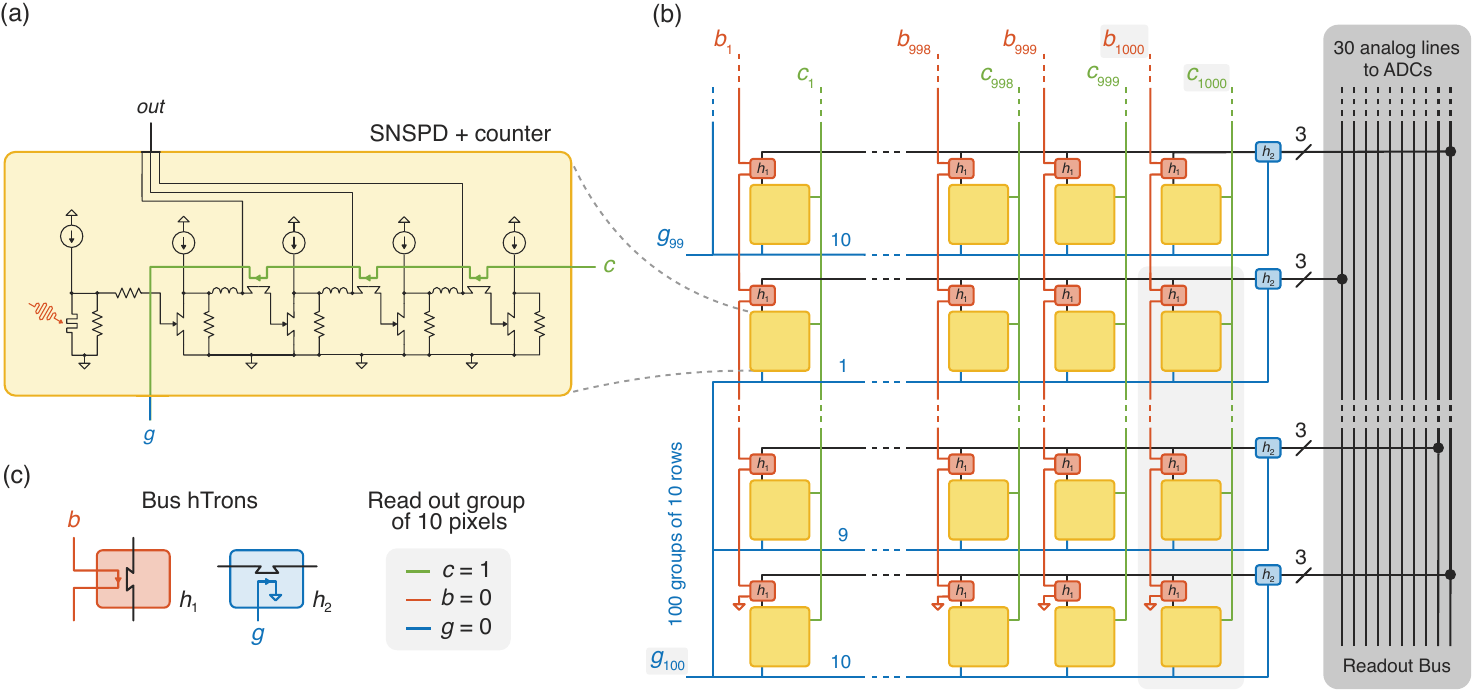}
    \caption{Design of a readout scheme for SNSPD megapixel using 3-digit counters. (a) Circuit schematic of a single pixel, which is composed of an SNSPD coupled to a 3-digit nanocryotron counter. The heaters of the hTrons in the counter are connected in series and the two terminals of the series are port $c$ and port $g$. If $c = 1$ (voltage high) and $g = 0$ (ground) the current flows into the heaters, so the output voltages are sent to the 3-digit $out$ port.
    (b) Schematic of a $1000 \times 1000$ SNSPD array.
    The array is divided into 100 groups of $10 \times 1000$ pixels. For each pixel, the $c$ port is connected to a column-selection line (green) capable of states 1 and Z (floating). The $g$ port is connected to a digital group-selection line (blue) capable of states 1 and 0. The states of 10 pixels are read by 30 3-bit analog-to-digital converters (ADCs) at a time. The latter are connected to the pixels through 3-digit vertical buses and 3-digit common row buses (black lines). (c) Connections between pixels and common row buses are controlled by hTrons $h_1$ (red). Their state (open or close switch) is set by the vertical $b$ lines. Connections between common row buses and the vertical buses of the ADCs are controlled by hTrons $h_2$. Their state is set by the $g$ lines. A group of 10 pixels is read out and connected to the ADCs if the corresponding selection lines are: $c = 0$, $b=0$, and $g=0$. If $c, g$ = Z,\,0; Z,\,1; 1,\,1 the counters are not read out. Selected pixels or control lines are highlighted in gray.\label{fig: array} }
    \end{figure*}
    As stated in section \ref{sec:intro}, the on-chip SNSPD-counter integration would be valuable for megapixel detector arrays that require high precision on the number of detection events. 
    In this section, we evaluate the potential capabilities of a megapixel-scale array, in which each pixel is formed by an SNSPD coupled to a multi-digit nanocryotron counter. In such a structure, the photon sensors are kept active for a chosen acquisition time, then disabled, and finally all the counters are read out. 
    As explained in section \ref{sec2:fab}, the SNSPD needs to be patterned with a film thickness on the order of 5\,nm to obtain saturated photon detection efficiency. Therefore, the counter needs to be redesigned for a thinner film or to be fabricated on a separate and thicker superconducting layer (see the discussion in section \ref{sec:dis}). 

    The number of digits and the base of the counters are chosen according to the expected count rate in the acquisition time interval. It is worth mentioning that by operating in base 2 rather than a base $N>2$, the output signal would already be binary. However, counting in $N>2$ would allow for fewer stages. Considering the large footprint of the current loop design ($70\,\text{µm}\times 50\,\text{µm}$), it is advantageous to operate the counter in the highest base possible to minimize the number of stages per pixel, in a megapixel-scale array. In this case, the output signal would be converted to binary through analog-to-digital converters.
    It is plausible to design an SNSPD megapixel array that relies on base-6 counters because: (1) we observed a $10^{-3}$ bit error rate that can be further decreased by optimizing the circuit for base 6 (see the discussion in section \ref{sec:dis}); and (2) we demonstrated by simulation parallel readout of a base-6 counter.

   As an example of architecture, we propose a $1000 \times 1000$ array, where each pixel includes a 3-digit base-6 counter with parallel readout. A single pixel can record up to 215 detection events before turning over. If each counter can be read out at a rate of 5000 reads per second such an array can accommodate up to $10^{12}$ detection events per second. Reading out a digit in a base-6 counter requires at least a 3-bit ADC. However, it is infeasible to provide an ADC for each digit of each counter in the array, thus a bus architecture is required. 
    
    Figure \ref{fig: array} shows the schematic of such an architecture. Each pixel (see figure \ref{fig: array}\textcolor{blue}{(a)}) includes a detector coupled to a 3-digit nanocryotron counter, with two control ports ($c$ and $g$ in the figure), and a 3-digit output $out$. The 3 heaters of the readout hTrons in the counter are connected in series. $g$ (row group) and $c$ (column) are the two terminals of the series of heaters. $g$ can be externally set to the high-impedance state (Z) or 1, and $g$  can be set to 0 or 1. When $c = 1$ and $g = 0$, a current is applied to the heaters so that the output voltages are generated. In all the other combinations of control signals, the counter is not read out. All the pixels have an additional shared control line (not shown in the figure) to turn off the bias of the SNSPDs after image acquisition.

    As shown in figure \ref{fig: array}\textcolor{blue}{(b)}, the array is divided into 100 groups of $10 \times 1000$ pixels, and they are selected one by one, during the readout procedure. In particular, each group is measured column by column, so that only 10 counters are simultaneously read out. 
    The $g$ ports of each group are controlled by a single group-selection line (blue in the figure), and the $c$ ports of each column are controlled by a single column-selection line (green in the figure). A column of 10 pixels is addressed if the corresponding column-selection line is high ($c=1$) and the group-selection line is low ($g=0$). 
    With this procedure, the entire array can be measured by 30 3-bit ADCs in 100\,000 steps. The ADCs must have a 2\,ns sample time to support tera-counts per second.
    
    A common horizontal bus system is necessary to make one million pixels communicate with the 30 ADCs connected to the vertical readout buses (grey box in figure \ref{fig: array}\textcolor{blue}{(b)}). The pixels of each row share a 3-digit horizontal bus, and only the 10 buses associated with the selected group are connected to the readout buses. Each of the 10 horizontal buses needs to connect to one pixel at a time. 
    These connections need to be controlled by switches to avoid cross-talk between pixels, which have a low output impedance. 
    Traditionally, an enhancement-mode MOSFET might be considered since they would provide a high-impedance channel in their quiescent state (i.e. they are normally-open switches). However, integrating cryo-CMOS devices into a high-density superconducting nanowire array remains a daunting challenge. One possible solution within the nanowire-electronics family is to use hTrons, which are normally-closed switches that ensure electrical isolation between the control ports and the channels. If a current passes through the heater (high heater signal), the switch is open. Otherwise, the switch is closed. Each pixel-to-horizontal-bus connection and each horizontal-bus-to-ADC connection is controlled by one hTron ($h_1$ and $h_2$ in figure \ref{fig: array}\textcolor{blue}{(b)} respectively). The heater signals of all the hTrons are normally high. When the buses and pixels are selected, the corresponding heater signals are low.
    Each group-selection line can be used to control both the $g$ ports and the heaters of the $h_2$ hTrons. 
    A single line $b$ (red in the figure) can control the heaters of all the $h_1$ in the same column. The state of this line, with additional logic elements, may be set by the signal on the corresponding column-selection line ($b=0$ when $c=1$, and $b=0$ when $c=$ Z).

    There are a total of 1100 control lines for the array readout: 1000 column-selection lines and 100 group-selection lines. Two possible architectures can be used to reduce the number of control lines further: (1) 10-to-1000 and 7-to-100 decoders, possibly based on the nanocryotron logic family \cite{buzzi_nanocryotron_2023}, address columns and groups respectively; additionally, two binary nanocryotron counters set the column-group address to progressively read out all the 10-pixel groups, using just two clock signals (one per counter, as input signal); 
    (2) 1000 bit and 100 bit nanocryotron shift registers \cite{foster_superconducting_2023}, which operate as one-hot counters, address columns, and groups respectively; each cell of the shift registers drives the heaters of bus hTrons, on a single column or group selection line; this configuration needs four clock signals (two out-of-phase clocks for each shift register). 
    In both architectures, the nanocryotron circuits need to generate the high impedance state (Z) for the unselected columns. The state can be provided by an additional hTron on each column-selection line that stops the current flow through it when activated. 

    The ADCs can be based on room-temperature electronics or cryo-CMOS systems working at higher-temperature stages of the cryostat. Increasing the number of ADCs would favor a parallel measurement, thus decreasing the readout time and the number of control lines. The optimal trade-off is application-dependent. 
    Assuming that the bus hTrons were designed to cool down, thus reset, in less than the 2\,ns sample time of the ADCs, the entire array would be read out in 200\,µs. Therefore, the collection of $10^{12}$ detection events per second would be achievable.
    
    Most of the power consumption of this architecture comes from the need to have current through the heaters in the majority of bus hTrons. There are always 30 heaters on for the selected pixels, $1000 \times 999$ heaters on for the pixels not selected, and 990 heaters on for the rows not addressed. Assuming the power consumption per hTron was on the order of 100\,nW (likely achievable by dimensional scaling from existing prototype demonstrations at $\sim 2$\,µW \cite{butters_digital_2022}), the total power consumption would be approximately 100\,mW. This value is compatible with the typical cooling power of 4\,K cryostats. The energy consumption of a heater should be studied experimentally to improve these estimates.
    The estimation does not include the power dissipated by the channels of nTrons and hTrons, which should be orders of magnitude lower than 100\,mW, since the channels are not constantly biased. Also, the power consumed by additional electronics for column-group addressing is not considered. This value would be much lower than 100\,mW if nanocryotron circuits were used. The static power dissipated by the bias resistors is not taken into account. 
    The latter could be reduced to negligible values by implementing biasing techniques used in RSFQ platforms, such as introducing superconducting inductors in series with bias resistors to reduce the required resistance for a sufficiently large output impedance \cite{kirichenko_zero_2011}. As a trade-off, this scheme would increase the pixel footprint. 
    
    Further device design and testing will be necessary to confirm the assumptions made in this analysis and to compare the solution with alternative array architectures.
    A potential drawback of this system is the heat dissipation of bus hTrons, which might cause a local temperature increase of pixel elements and thus alter their bias margins. In the future, such temperature changes in counters and SNSPDs should be estimated in heat-transfer simulations and the circuit design should be updated accordingly to consider a decrease in the critical currents.

\section{\label{sec:dis} Discussion}
    This work is a proof-of-concept demonstration of a superconducting nanowire ripple counter. We showed that it is possible to perform sensing and logic operations in a single superconducting technology. 
    However, further work is needed to experimentally confirm the design of either the parallel or serial readout scheme. Moreover, the performance of the circuit should be optimized, such as by reducing the device footprint or operating at higher speeds.

    An important step towards complete integration with SNSPDs would be to use on-chip resistors instead of surface-mount components, to shunt the nanowires and to couple the SNSPD to the counter. Such resistors can be fabricated with titanium-gold thin films using a multi-layer process \cite{castellani_design_2020}.
    
    In the current design, the chip area occupied by each loop in the counter is dominated by the kinetic inductor $L_0$, as shown in figure \ref{fig: 2bit}\textcolor{blue}{(b)} ($40\,\text{µm}\times 40\,\text{µm}$). Reducing the footprint of the inductor is fundamental for scaling up the technology. We chose $L_0$ according to the behaviors observed in simulations and we did not minimize its value with more detailed experimental analysis. As explained in previous sections, the inductor is critical for the correct functioning of each stage. It lets only a small portion of $I_\text{b1}$ flow into the loop, keeping $nT_1$ at the correct bias point. More precisely, the higher the ratio between the input impedance of the loop and the output impedance of the shunted $nT_1$, the lower the value of $\Delta i$. The output impedance depends on the drain inductance and the shunt resistance of $nT_1$. The input impedance of the loop is dominated by $L_0$.
    It is possible to decrease the dimension of $L_0$ if a high input-to-output impedance ratio (e.g. $>10$) is preserved. Therefore, removing the drain inductors $L_\text{d}$ would allow lower $L_0$. In this study, the drain inductance was included to avoid the latching of the nTron channel observed in LTspice simulations \cite{Note1}. However, we did not study experimentally how much $L_\text{d}$ could be decreased before observing latching. Other measurements of shunted nanowires suggested that the LTspice model overestimates the latching effect, meaning that a smaller drain inductance might be achievable.  
    
    Another method to reduce the size of the inductor would be decreasing the shunt resistances to lower the output impedance of $nT_1$. However, this approach would result in a smaller output signal. 
    It is worthwhile to note that the inductor size could be scaled downward by scaling down the width of all the nanowires in the circuit, including the nTron gate and channel widths. However, with our current fabrication methods, scaling the gate widths significantly below 30\,nm reduces the reproducibility and reliability of the circuit.

    In the best-case scenario, the loop inductance of the demonstrated device could be scaled down by applying all the methods described above, still keeping the same range of base tunability. According to our estimation, removing the drain inductors, scaling down the widths of all the nanowires by a factor of 3/4, and lowering the shunt resistances by a factor of 3, would allow decreasing the value of $L_0$ from 140\,nH to about 30\,nH. Therefore, the footprint would scale down from $40\,\text{µm}\times 40\,\text{µm}$ to approximately $15\,\text{µm}\times 15\,\text{µm}$. 
    
    The measured bias margins are large enough for the counter to operate stably and correctly at rates up to $10^7\ s^{-1}$. 
    Flood-illuminating both counters and detectors on the same chip did not substantially alter the performance of the system. The reason is that the loop inductors, which dominated the footprint of the counter, were always biased out of the saturation regime of detection efficiency ($i_\text{L}\ll I_\text{C}$). The only regions in which detection events were more likely to occur were nTron chokes and channels, which occupied a very small fraction of the total area. Therefore, the probability of bit errors due to photons was low. The problem should be studied by applying more input pulses for a longer time. If photon detection in the counter turns out to be an issue, the counter will need to be shielded from light. More generally, the counter should be characterized with more clock pulses to verify if a bit error rate lower than $10^{-4}$ can be achieved.

    Also, the presence of an out-of-plane magnetic field reduced the bias margins. This effect on loops containing nanocryotrons has been analyzed more rigorously by \cite{buzzi_nanocryotron_2023} and \cite{foster_superconducting_2023}. In general, nanocryotron-based circuits are more resistant to flux noise than SFQ architectures, mainly because the loops store several fluxons (on the order of 100) instead of one. 
    The circulating current due to the geometric inductance of the loop is induced only when the magnetic field is ramped up. Therefore, it vanishes after the first hotspot is formed in the loop. 
    On the other end, the Meissner effect could be the main cause of errors in nanocryotron circuits. The effect, combined with current crowding \cite{hortensius_critical-current_2012}, favors vortex penetration into the film, in the proximity of sharp edges of the nTron choke. The vortex penetration may generate undesired hotspots in the loops, thus bit errors. 
    
    We designed the stages to have a tunable counting base.
    However, we could show tunability only up to base 6, and we observed a significant BER, particularly in the higher bases (BER = $10^{-3}$ at base 6, 5\,MHz). The range of possible $N$ might be enlarged by designing the nTron geometry to maximize critical-current suppression actuated by the gate to the channel. In fact, larger critical-current suppression would allow biasing the nTron at a lower $I_\text{b}$, thus increasing the number of allowable current levels in the loop. 
    If tunability was not relevant to a particular application, the counter design could be optimized for a single counting base, maximizing the bias margins and minimizing the circuit footprint.
    For example, in base 2 the ratio between channel and gate widths could be decreased so that the bias current of each nTron would be as close as possible to the gate critical current of the next nTron. By doing so, the loop inductance, and hence the circuit footprint, could be reduced.
    Also to obtain better bias margins in base 6, the ratio between channel and gate widths would be decreased but the loop inductance would remain the same to allow more current levels in the loop. 
    
    Broadening the bias margins is essential to making the system resistant to fabrication variations. In fact, different gate widths may require different $I_\text{b}$ for each nTron. If the margins are large enough all the nTrons can be biased with the same value, simplifying the structure. In this case, a single line with a resistive or inductive bias distribution tree could control the bias levels of the entire system \cite{kirichenko_zero_2011}. 
    
    The maximum count rate is limited by the reset time of the nTrons and the time constant of the loop inductor. Decreasing the loop inductance would improve the speed of the system. The maximum operating frequency is intrinsically limited by the thermal time constant of NbN nanowires \cite{kerman_electrothermal_2009}. However, we note that SNSPDs and counters share the same limitation, and it might be possible to optimize the circuit to make the counter's recovery time lower than that of a reasonably sized SNSPD.  
    In the future, the margins of the counter should be studied at count rates higher than $10^7\,$s$^{-1}$ (the value demonstrated in this work), using the same approach of \cite{buzzi_nanocryotron_2023}. Moreover, it would be interesting to directly measure the flux in the loops (e.g. using a SQUID: superconducting quantum interference device), so that all our hypotheses on bit-errors causes would be verified. Also, the retention time of the states in the loops should be studied rigorously.

    In section \ref{sec:array}, we discussed designing a megapixel-scale SNSPD array. 
    The current circuit design would allow having a minimum pixel footprint on the order of $100\,\text{µm}\times 100\,\text{µm}$, for a maximum photon count of 124. The structure would include an SNSPD and a 3-digit counter working in base 5. Each loop would occupy approximately $50\,\text{µm}\times 50\,\text{µm}$, and the SNSPD area would be equal or lower than $50\,\text{µm}\times 50\,\text{µm}$. 
    If the loop inductor footprint could be scaled down to $15\,\text{µm}\times 15\,\text{µm}$ as mentioned above, the minimum pixel area would be on the order of $30\,\text{µm}\times 30\,\text{µm}$, with a $15\,\text{µm}\times 15\,\text{µm}$ SNSPD. In this scenario, the maximum array fill factor would be about 25\,\%. 
    The real footprint might be larger if the inductive bias distribution tree was implemented to lower the overall power consumption. Therefore, for the same SNSPD dimensions, the array fill factor could be even lower.

    Such values of footprint and fill factor would limit the creation of a megapixel array. Moreover, patterning detectors and electronics on the same superconducting film is not optimal for simultaneously optimizing pixel area, margins, and photon detection efficiency. The development of a multi-level process would help scale down the pixel dimensions, increase the fill factor, and improve the detector performance. A possible solution might be the following: the SNSPD is on the top NbN layer and the stages of the counter are distributed in the underlying ones (one per layer, including the bias inductor) so that the detector occupies the entire pixel area; the levels are separated by a dielectric layer, and interconnected through superconducting vias. Additionally, a metallic layer is used for the shunt resistors.  Another valuable solution would be using multi-layer hTrons \cite{baghdadi_multilayered_2020} instead of nTrons so that the heaters would heat up the channels from one layer to another without the need for vias. The hTrons would need to have superconducting heaters because a resistive heater would continuously discharge the loops erasing the stored states.

    The reproducibility of counters with large and consistent bias margins should be further studied to evaluate the feasibility of designing megapixel arrays. We observed correct counting operations with sporadic bit errors in multiple devices on different chips. However, we did not accurately estimate the BER for all counters, so additional characterizations are required to confirm that devices with the same or similar geometries have comparable bias margins.  

    The proposed array architecture has some drawbacks in comparison with the existing solutions \cite{oripov_superconducting_2023}. First of all, it does not exploit the high temporal resolution of SNSPDs to record the photon arrival times at each pixel. However, additional readout circuitry can be designed to preserve the time information of specific pixels, while counters are operating (e.g. adding readout lines thermally coupled to multiple detectors as in \cite{oripov_superconducting_2023}). A second limiting factor for the development of this architecture is the need to develop a multi-layer process to maximize the fill factor and detection efficiency. 
    On the other end, one of the main advantages of this architecture is the high accuracy of the number of photons. Indeed, parallelizing the counting operation at the pixel level ensures that all the detected photons are counted, and using SNSPDs guarantees extremely low dark count rates. In addition, the number of photons is digital, simplifying the communication with superconducting electronics for further signal processing at low temperatures. The drawback of digitization is that a single error on the most significant digit of the counter can cause a large change in the stored number.

\section{\label{sec:conc} Conclusion}

    In summary, we presented an architecture for a superconducting ripple counter based on nanocryotrons that can perform signal processing, encoding, and storing on chip for the output of SNSPDs. We demonstrated the fundamental circuit component of the counter, showing an analogy with the logical behavior of a T flip-flop, or frequency divider. Moreover, we showed that the divisor $N$ of the frequency divider can be tuned from 2 to 5. 
    We experimentally verified that the integer-$N$ frequency divider could be used to create a 2-digit base-$N$ ripple counter. This circuit counted the number of photons detected by an SNSPD in modulo 4 or 9 (i.e. base 2 or 3) with a low bit error rate.
    We proposed serial and parallel schemes, demonstrated by simulations, to read out the state of the counter, and we analyzed a possible readout system for an SNSPD megapixel array that can record up to tera-counts per second. 
    
    These results will help develop the next generation of superconducting detector arrays, which rely on cryogenic edge computing. Edge computing will be fundamental to scale up SNSPD systems for applications such as: (1) mid-infrared imaging in low-illumination environments; (2) optical communication in deep space \cite{shaw_superconducting_2017}; (2) particle detection in large high-energy-physics experiments \cite{polakovic2020unconventional}; and (3) photonic Bell-state measurements for arrays of spin memories in quantum repeaters \cite{bhaskar_experimental_2020}.

    In a more general perspective, the demonstration of the ripple counter, combined with existing examples of nanocryotron components (e.g. logic gates, and shift register), will boost the development of low-energy superconducting processors.

\section{Acknowledgments}
    The initial stages of the research were sponsored by the Army Research Office (ARO) and were accomplished under Cooperative Agreement Number W911NF-21-2-0041. Support was provided by the Defense Advanced Research Projects Agency (DARPA) Defense Sciences Office (DSO) Invisible Headlights program. The views and conclusions contained in this document are those of the authors and should not be interpreted as representing the official policies, either expressed or implied, of the Army Research Office or the U.S. Government. The U.S. Government is authorized to reproduce and distribute reprints for Government purposes notwithstanding any copyright notation herein.
    The completion of data analysis and presentation was sponsored by the Center for Quantum Networks (CQN) under the National Science Foundation (NSF) grant number 1941583.
    O.M. acknowledges support from the NDSEG Fellowship program. 
    M. Colangelo acknowledges support from MIT Claude E. Shannon award. 
    A.B. acknowledges support from Politecnico di Torino.
    R.F. acknowledges support from the DOE under the National Laboratory LAB 21-2491 Microelectronics grant.
    The authors would like to thank Kyle Richards and Teja Kothamasu for their assistance with setting up and using the Keysight PXIe system. The data that support the findings of this study are available from the corresponding author upon reasonable request. The authors have no conflicts of interest to report.
    The authors would like to thank Francesca Incalza for helping with the characterization of the detector. 
    The authors would like to thank Stewart A. Koppell and Phillip D. Keathley for helping review the manuscript.

\bibliography{main}

\end{document}


\supplementarysection
\preprint{APS/123-QED}

\title{Supplemental Material\\ \small Nanocryotron ripple counter integrated with a superconducting nanowire single-photon detector for megapixel arrays}



\author{Matteo Castellani}
\email{mcaste@mit.edu}
\affiliation{%
 Electrical Engineering and Computer Science, \\ Massachusetts Institute of Technology, Cambridge, Massachusetts, United States
}%
\author{Owen Medeiros}%
\affiliation{%
 Electrical Engineering and Computer Science, \\ Massachusetts Institute of Technology, Cambridge, Massachusetts, United States
}%
\author{Reed A. Foster}%
\affiliation{%
 Electrical Engineering and Computer Science, \\ Massachusetts Institute of Technology, Cambridge, Massachusetts, United States
}%
\author{Alessandro Buzzi}%
\affiliation{%
 Electrical Engineering and Computer Science, \\ Massachusetts Institute of Technology, Cambridge, Massachusetts, United States
}%
\author{Marco Colangelo}%
\affiliation{%
 Electrical Engineering and Computer Science, \\ Massachusetts Institute of Technology, Cambridge, Massachusetts, United States
}%
\author{Joshua C. Bienfang}%
\affiliation{%
Joint Quantum Institute, National Institute of Standards and Technology and University of Maryland, 100 Bureau Drive, Gaithersburg, Maryland 20899, USA}%

\author{Alessandro Restelli}%
\affiliation{%
Joint Quantum Institute, National Institute of Standards and Technology and University of Maryland, 100 Bureau Drive, Gaithersburg, Maryland 20899, USA}%

\author{Karl K. Berggren}%
\affiliation{%
 Electrical Engineering and Computer Science, \\ Massachusetts Institute of Technology, Cambridge, Massachusetts, United States
}%




\date{May 9, 2024}

\maketitle


\newpage

\section{\label{sec:field} Bit error rate with an applied magnetic field}

    \begin{figure*}[htbp!]
    \includegraphics[width=8cm]{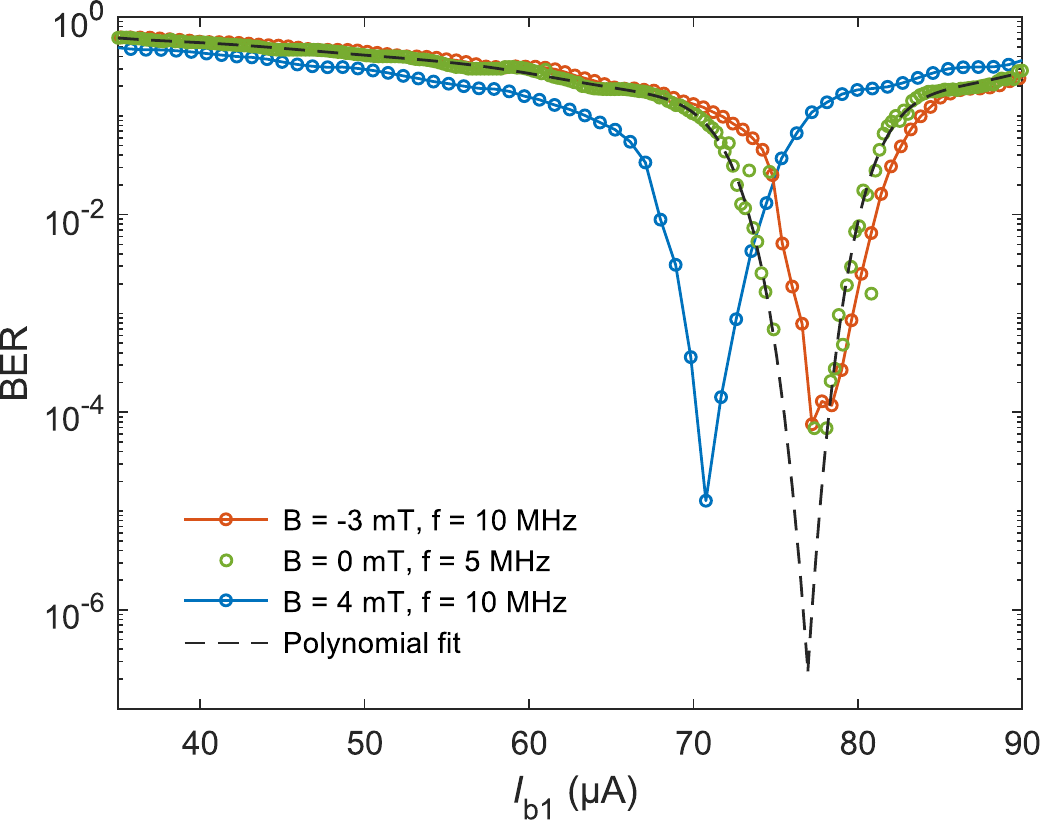}
    \caption{Experimental bit error rate (BER) of a single stage of the counter operating in base 2 as a function of the bias current $I_\text{b1}$, for two different values of the out-of-plane magnetic field with 10\,MHz input frequency (clock), compared to the BER at zero magnetic field and 5\,MHz input frequency. $I_\text{b2} = 85$\,µA and $i^\text{max}_\text{IN} = 72$\,µA (input peak current) were kept constant. The black dashed curve is a composition of two polynomial fits of the bit error rate with a zero applied field (outliers were excluded from the fit). The figure was obtained from device C01 on SPG687. \label{fig: ber_field}}
    \end{figure*}

    We characterized a stage of the counter with an applied out-of-plane magnetic field to study how the BER changes. Figure \ref{fig: ber_field} shows the BER in base 2 as a function of $I_\text{b1}$ with two values of the magnetic field, in comparison with the data of figure \textcolor{blue}{3(b)}, obtained without a field. Without a magnetic field, we estimated a minimum BER of about $3 \times 10^{-7}$ by applying a polynomial fit to the data of figure \textcolor{blue}{3(b)}. The data with the field were obtained with a 10\,MHz input frequency, while the data at zero field were obtained at both 5\,MHz and 10\,MHz. We could not estimate the BER at 10\,MHz without field because the measurement was altered by high background noise. Therefore, we could not compare the BER with a field and without a field, at 10\,MHz. However, we could conclude that the effects depended on the field orientation. In particular, a negative field induced a larger increase of the BER, with respect to a positive field. This might depend on several factors, such as the direction of circulating currents in the circuit and the asymmetric geometry of nTrons. With $B = -3$\,mT and $f = 10$\,MHz, the margins shrank down by about 50\% if compared with the data at $B = 0$\,mT and $f = 5$\,MHz, and considering $10^{-4}$ as the maximum tolerable BER.

\newpage

\section{\label{sec:char} Characterization of the SNSPD}

   \begin{figure*}[htbp!]

    \includegraphics[width=16cm]{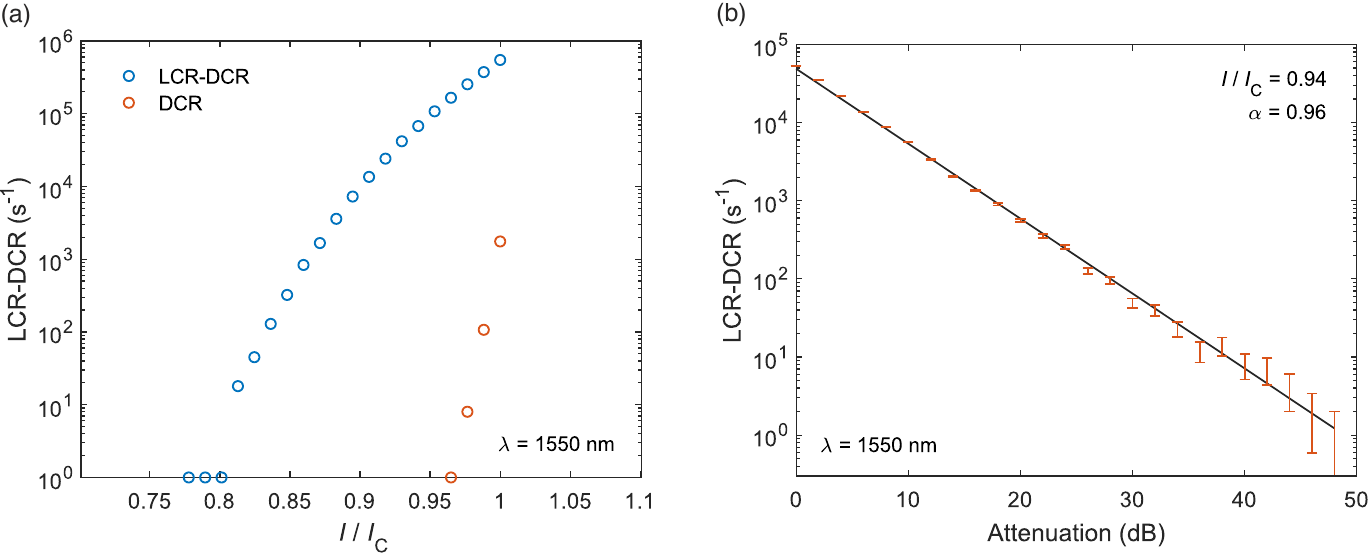}
    \caption{Characterization of the SNSPD used to demonstrate photon counting. (a) Light counts rate minus dark counts rate (LCR-DCR) and dark counts rate (DCR) as a function of the detector bias current (normalized by the critical current). $\lambda = 1550$\,nm. (b) Light counts rate minus dark counts rate as a function of the laser power attenuation at $\lambda = 1550$\,nm. Error bars indicate the standard deviation for the Poisson distribution associated with each data point. The black curve is the linear fit of the experimental data. The fitting coefficient is $\alpha = 0.96$, if the count rate is expressed in dB. The detector is biased at $I/I_\text{C}=0.94$. For both plots, the count rates were obtained with an integration time of 1 second. Nominal laser power without attenuation: $P = 2.1$\,mW.\label{fig: char}}
    \end{figure*}

We characterized the SNSPD used to demonstrate photon counting by flood illuminating it with 1550\,nm light. The optical coupling between the fiber and the detector was weak. Indeed, we estimated an extremely low quantum efficiency on the order of $10^{-6}$.
Figure \ref{fig: char}\textcolor{blue}{(a)} shows the photon count rate (PCR) as a function of the detector bias current. The device could count up to about $10^{6}$\,$s^{-1}$ but the PCR did not saturate. Probably, the saturation could not be achieved because the superconducting film was too thick. Using a thinner film (e.g. 5\,nm) would solve this problem but the output current of the detector would decrease. Therefore, the first nTron of the counter would need to be redesigned to sense smaller signals.

Figure \ref{fig: char}\textcolor{blue}{(b)} shows the PCR as a function of the laser power attenuation, with fixed bias current. Considering that the slope of the curve is $\alpha = 0.96$, linearity and thus single photon detection were verified. 
This result obtained at 1550\,nm implies that the detector worked in the single photon detection regime also at higher photon energies, and thus at 405\,nm. Therefore, we could confirm that the digital number stored in the counter was associated with an actual number of detected photons.

\section{Photon counting at 1550\,nm}

   \begin{figure*}[htbp!]
    \includegraphics[width=8cm]{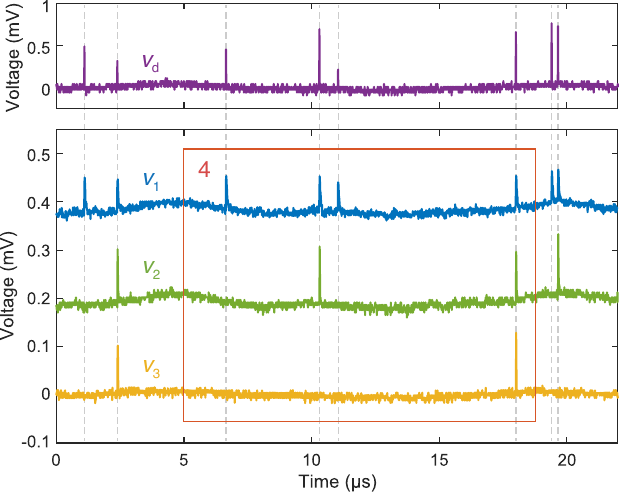}
    \caption{Experimental time-domain behavior in base 2 of a superconducting nanowire 2-digit counter coupled to an SNSPD, flood illuminated at 1550\,nm. Colors and labels associate the waveforms with currents and voltages in the schematic of figure \textcolor{blue}{5(a)}. The voltage traces were acquired after the amplification. In the figures, voltage amplitudes are divided by the gain of the amplifiers at 1\,GHz reported in the datasheet (45 dB). The red box highlights the count of 4 pulses. The waveforms are vertically shifted for clarity. The data were obtained with the same device used for figure \textcolor{blue}{5(a)}. All acquired traces have low-frequency noise in the background. \label{fig: counts1550}}
    \end{figure*}

We demonstrated correct photon counting operation in base 2 at 1550 nm, with the same experimental setup used to count at 405\,nm. Figure \ref{fig: counts1550} shows the experimental time-domain response of the circuit. We did not estimate the bias margins of the system at 1550\,nm due to the lower photon count rate with respect to the experiment at 405\,nm. However, the BER should not depend on the wavelength of the detected photons. Therefore, the values of BER at 405\,nm and 1550\,nm would likely be comparable. 

\section{\label{sec:ser} Serial readout}

   \begin{figure*}[htbp!]
    \includegraphics[width=16cm]{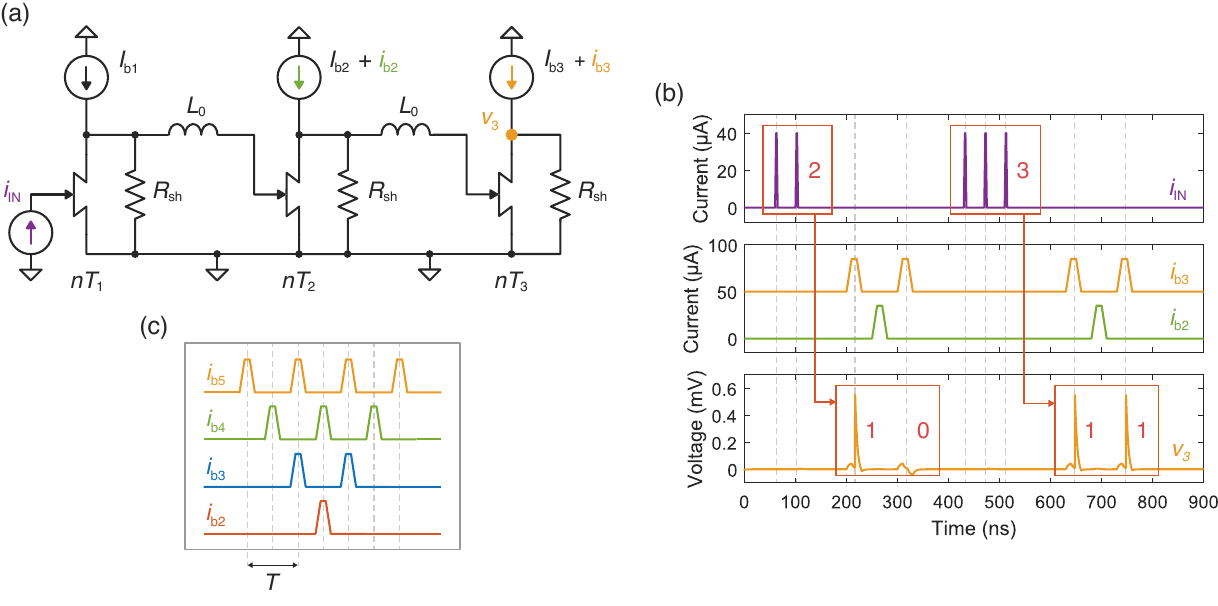}
    \caption{Design and simulation of a serial-readout method for a 2-digit counter in base 2. (a) Circuit schematic of the readout scheme. The structure is the same as a standard nTron 2-digit counter without additional components. The $i_\text{b2}$ and $i_\text{b3}$ signals are sent to the bias lines to shift the states of the two stages towards the loop of the most significant digit (the right-most loop). $I_\text{b1}$, $I_\text{b2}$, and $I_\text{b3}$ are kept constant. The output voltage is read on the $v_3$ node ($Q^\text{out}_\text{n}$). (b) Simulated time-domain characteristics of the circuit. Two input pulses are counted and read out (and automatically reset). Then, three pulses are counted and read out. Readout process: a leading current pulse ($i_\text{b3}$) is sent to the channel of $nT_{3}$ to read the second digit. If there is a non-zero current in the second loop ($Q_2 =1$) the nTron generates an output voltage pulse at $V_3$. Then, the first-loop state is shifted to the second loop by a secondary current pulse on $i_\text{b2}$. Afterward, the output for the first digit is generated by applying another, final pulse on $i_\text{b3}$. The red boxes show the digital state on the output line. $Q=1$ is encoded in the presence of a spike, $Q=0$ in the absence of a spike. This type of serial readout has been simulated with up to 5 input pulses. Waveforms for 0, 1, 4, and 5 pulses are not shown here for simplicity. The colors and labels associate the waveforms with currents and voltages in the schematic. (c) Example of readout clock signals for a 4-digit counter. $i_\text{b5}$ is associated with the last nTron ($nT_5$), $i_\text{b2}$ with the second nTron ($nT_2$). Each signal $i_\text{bn}$ other than $i_\text{b5}$ is delayed by $T/2$ (half period) with respect to $i_\text{b(n+1)}$. \label{fig: serial}}
    \end{figure*}

\pagebreak
   Figure \ref{fig: serial} shows the serial readout scheme for a 2-digit counter in base 2. As opposed to the parallel readout, this system does not need an hTron for each loop. The states of the loops are shifted to the last cell by the application of out-of-phase clock pulses on the bias lines, and read out as a stream of voltage spikes across the last nTron ($nT_3$ in the figure). 
    The readout procedure starts from the most significant digit and ends with the least significant. First, a 35\,µA pulse ($i_\text{b3}$) is added to the 180\,µA bias current ($I_\text{b3}$) of the last nTron, which switches only when the sum of the gate current  $i_\text{g3}$ and total channel current $I_\text{b3} + i_\text{b3}$ exceeds the channel threshold. Therefore, if the state of the last loop is 1 an output voltage spike is generated on $v_3$. If the state is 0 the output voltage stays at zero. Afterward, the first digit is shifted to the second loop by applying the same procedure to $nT_2$, and the second output spike is formed with another pulse on $nT_3$.
    
    The scheme can be extended to $N$-digit counters, with a readout time that grows as $N$. Shifting each digit to the adjacent loop cannot be done simultaneously, otherwise, the states would be altered. Therefore, it is necessary to use $N$ different clock signals $i_\text{bk}$ ($2<k<N+1$) with period $T$ and progressive delay $(N+1-k)T/2$ from the clock of the last nTron $i_\text{b(N+1)}$. Thus, the shift to the adjacent loop is performed simultaneously only for some of the odd or even loops alternately, without modifying the states.
    Figure \ref{fig: serial}c shows the clock signals for reading out a 4-digit counter. 
    The need for $N$ different clock signals is the main limitation for scaling this approach to larger counters. A possible solution may be to combine the structures of the counter and the nTron shift register recently demonstrated by [16]. In this approach, an auxiliary loop between all adjacent cells is introduced so that it is possible to have just two out-of-phase clock signals for any value of $N$, without altering the stored states. 

    The advantages of the serial readout scheme relative to parallel readout are: (1) a simpler circuit structure; and (2) the need for only a single output line. 
    The disadvantages are: (1) it can be implemented only in base 2; (2) the readout time is $N$ times longer; and (3) $N$ different control signals are needed.